\documentclass[prb,twocolumn,superscriptaddress]{revtex4}
\usepackage{amsmath}
\usepackage{amsfonts}
\usepackage[percent]{overpic}

\begin{document}
\title{Entangled-state generation and Bell inequality violations in
nanomechanical resonators}
\author{J.~Robert Johansson}
\affiliation{iTHES Research Group, RIKEN, Saitama 351-0198, Japan}
\author{Neill Lambert}
\affiliation{CEMS, RIKEN, Saitama 351-0198, Japan}
\author{Imran Mahboob}
\affiliation{NTT Basic Research Laboratories, Nippon Telegraph and Telephone Corporation, Atsugi 243-0198, Japan}
\author{Hiroshi Yamaguchi}
\affiliation{NTT Basic Research Laboratories, Nippon Telegraph and Telephone Corporation, Atsugi 243-0198, Japan}
\author{Franco Nori}
\affiliation{CEMS, RIKEN, Saitama 351-0198, Japan}
\affiliation{Physics Department, University of Michigan, Ann Arbor, Michigan, 48109, USA}

\date{\today}

\begin{abstract}
We investigate theoretically the conditions under which a multi-mode nanomechanical resonator,
operated as a purely mechanical parametric oscillator, can be driven into highly nonclassical states.
We find that when the device can be cooled to
near its ground state, and certain mode matching conditions are satisfied, it is possible to
prepare distinct resonator modes in quantum entangled states that violate Bell
inequalities with homodyne quadrature measurements. We analyze the parameter regimes for such
Bell inequality violations, and while experimentally challenging, we believe that the realization of such states
lies within reach. This is a re-imagining of a quintessential quantum optics experiment by using phonons
that represent tangible mechanical vibrations.
\end{abstract}

\maketitle

\section{Introduction}

Reaching the quantum regime with mechanical resonators have been a long-standing goal in the
field of nanomechanics \cite{cleland:2002, blencowe:2004,poot:2012,aspelmeyer:2013}.
In recent experiments, such devices have been successfully cooled down to near their
quantum ground states \cite{oconnell:2010, teufel:2011:2, chan:2011}, and in the future may be used
for quantum metrology \cite{regal:2008},
as quantum transducers and couplers between hybrid quantum systems
\cite{sun:2006, wallquist:2009, safavi-naeini:2011, xiang:2013, bochmann:2013},
for quantum information processing \cite{rips:2013},
and for exploring the limits of quantum mechanics with macroscopic objects.
In many of these applications it is essential to both prepare the nanomechanical
system in highly nonclassical states and to unambiguously demonstrate the quantum
nature of the produced states.

Nonclassical states of harmonic resonators can be achieved by introducing
time-dependent parametric modulation \cite{tian:2008} or via nonlinearities.
The latter can be realized by a variety of techniques, for example by coupling to a superconducting qubit\cite{oconnell:2010},
coupling to additional optical cavity modes\cite{jahne:2009,stannigel:2011,wang:2013},
applying external nonlinear potentials\cite{rips:2013},
or via intrinsic mechanical nonlinearities in the resonator itself\cite{westra:2010,lulla:2012,khan:2013,yamaguchi:2013}.
Using such nonlinearites, specific modes of a nanomechanical resonator could potentially be prepared in
a rich variety of different nonclassical states,
such as quadrature squeezed states \cite{clerk:2008, jahne:2009, liao:2011,palomaki:2013,palomaki:2013:2,lemode:2013},
subpoissonian phonon distributions \cite{qian:2012, nation:2013, lorch:2012}, Fock states \cite{rips:2012},
and quantum superposition states \cite{tian:2005, voje:2012, oconnell:2010}. Quantum correlations and entanglement
between states of distinct oscillator modes could also be potentially generated,
typically taking the form of entangled phonon states and two-mode quadrature correlations
and squeezing \cite{fei:2007, cohen:2013, huatang:2013, xu:2013}.
Experimentally, nonlinear interactions between modes of nanomechanical resonators have
already been used\cite{junho:2010,massel:2011} for parametric amplification and noise squeezing.
Various schemes\cite{eichler:2011,rips:2012,voje:2013} have proposed using nonlinear dissipation 
processes to realizing steady state entanglement. In a similar direction, a recent proposal\cite{jacobs:2012}
looked at ways to couple different internal mechanical modes of a nanomechanical system via
ancillary optical cavities. Also, Rips {\it et al.} [\onlinecite{rips:2012}] looked at ways to prepare nonclassical states
using enhanced intrinsic mechanical nonlinearities.

\begin{figure}[b]
\includegraphics[width=\columnwidth]{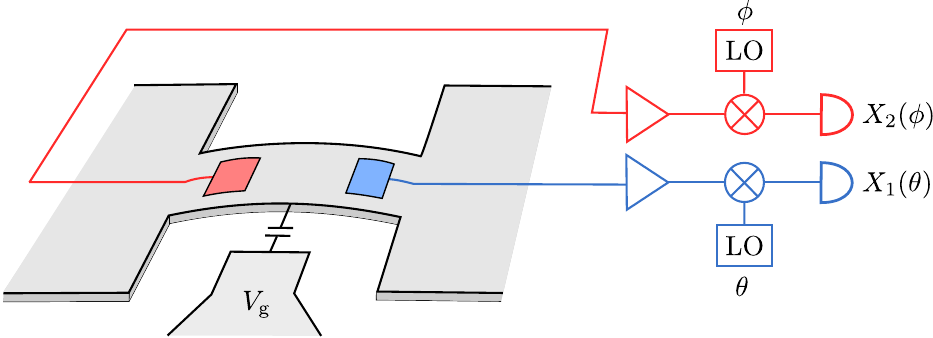}
\caption{
(Color online) Schematic illustration of a conceptual nanomechanical resonator
with two homodyne measurement setups that probe two different modes of oscillation.
The beam can oscillate in a large number of vibrational and flexural modes with different
frequencies. The two homodyne detectors measure the mode quadratures $X_1(\phi)$
and $X_2(\theta)$ with frequencies $\omega_1/2\pi$ and $\omega_2/2\pi$, respectively.
By analyzing the correlations between the $X_1(\phi)$ and $X_2(\theta)$ quadratures,
it is possible to determine whether or not the mode states are quantum-mechanically
entangled.}
\label{fig:schematic}
\end{figure}

Here we consider the generation of nonclassical states and the subsequent violation of Bell inequalities by the use of similar intrinsic mechanical
nonlinearities\cite{yamaguchi:2013, lulla:2012, khan:2013, rips:2013}.
We focus on a model relevant to a recent experimental realization \cite{mahboob:2013} of a phonon laser,
where a single mechanical device exhibits significant coupling between three internal modes of deformation,
due to asymmetries in the beam\cite{yamaguchi:2013}, and selective activation using external driving.
Here we examine that same intrinsic inter-mode interaction in the quantum limit. A schematic illustration of the device considered here is shown in Fig.~\ref{fig:schematic}, though this is not intended to be representative of the ideal realization or measurement scheme for operating in the quantum limit.  In most of our discussion we do not consider an explicit physical setup but rather focus on setting bounds on the fundamental system parameters necessary to realize the phenomena we discuss. The model we derive consists of an adiabatically-eliminated pump mode which drives the interaction between two lower-frequency signal and idler modes. We show that in the transient regime one can obtain violations of a Bell inequality based on correlations between quadrature measurements of the signal and idler modes. This is a re-imagining of a quintessential quantum optics experiment by using phonons that represent tangible mechanical vibrations. 

This paper is organized as follows: In Sec.~\ref{sec:model} we
introduce the general model and the Hamiltonian for a nonlinear nanomechanical device. In
Sec.~\ref{sec:paramp-regime} we consider a regime in which a parametric
oscillator is realized using three modes in the mechanical system, and in
Sec.~\ref{sec:paramp-regime-2mode} we introduce an effective two-mode model, valid when
the pump mode can be adiabatically eliminated, and we analyze the types of nonclassical states that can be generated in this system.
In Sec.~\ref{sec:quadrature-bell}, we review Bell's inequality using quadrature
measurements, and in Sec.~\ref{sec:mechanical-bell-violation} we analyze the
conditions for realizing a violation of this quadrature-based Bell inequality
with the mechanical system in the parametric oscillator regime studied in
Sec.~\ref{sec:paramp-regime-2mode}.
Finally we discuss the outlook for an experimental implementation using either intrinsic nonlinearities in
Sec.~\ref{sec:expr-outlook}, or, as an alternative, optomechanical nonlinearities in Sec.~\ref{sec:expr-outlook2}.
We summarize our results in Sec.~\ref{sec:conclusion}.

\section{Model}\label{sec:model}

The general Hamiltonian for a nonlinear multimode resonator,
describing both the self-nonlinearities and multimode couplings up to fourth-order,
can be written as \cite{khan:2013}
\begin{eqnarray}
\label{eq:full_hamiltonian}
H &=& \sum_k \omega_k a^{\dagger}_k a_k
+ \sum_{klm}\mathcal{\beta}_{klm}x_k x_l x_m \nonumber\\
&+& \sum_{klmn}\mathcal{\eta}_{klmn}x_k x_l x_m x_n + \mathcal{O}(x^5),
\end{eqnarray}
where $\omega_k$ is the frequency, $a_k$ is the annihilation operator, and
$x_k = a_k + a_k^\dagger$ is the quadrature of the mechanical mode $k$. Here the
basis has been chosen so that linear two-mode coupling terms are eliminated.
The third-order mode-coupling tensor $\mathcal{\beta}_{klm}$
describes the odd-term self-nonlinearity and the trilinear
multimode interaction. The fourth-order terms describes the even-term self-nonlinearity
and fourth-order multimode coupling. In symmetric systems the fourth-order terms dominate (odd terms vanish due to symmetry),
and it has been proposed elsewhere that they can be used to create effective mechanical qubits\cite{rips:2013}.
The possible combination of both third and fourth-order terms will be briefly considered in the final
section. The strength of the nonlinearity depends on the fundamental frequency
(length) of the resonator, and can be enhanced by a range of
techniques\cite{rips:2013}. In this work we focus on the three-mode coupling terms, as these  are necessary to generate the states that violate continous variable Bell inequalities.  Such terms vanish in symmetric systems and thus depend on the degree of asymmetry in the mechanical device\cite{khan:2013, yamaguchi:2013}, which again can be enhanced with fabrication techniques.  Our approach in the following is to identify the ideal situation under which one can realize these rare Bell inequality violating states.  Ultimately these states will be degraded by losses (which we investigate), but also by unwanted nonlinearities from the above hamiltonian.

\subsection{Parametric oscillator regime}\label{sec:paramp-regime}

\begin{figure}[t]
\includegraphics[width=\columnwidth]{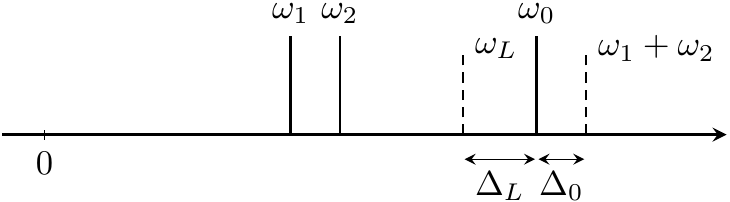}
\label{fig:modematching}
\caption{A visualization of the mode-matching condition required to obtain an effective three-mode system.
The modes are represented as solid vertical lines, and the driving frequency and the sum of the signal and
idler frequencies, which should sum up to a frequency close to $\omega_0$, are represented by dashed vertical
lines.}
\end{figure}

Nanomechanical devices of the type described in the previous section have a large number of modes
with different frequencies which depend on the microscopic structural properties of the beam. Here we focus on three such modes (relabelled as $k=0,1,2$) which are chosen such that they satisfy the phase-matching
condition $\omega_1 + \omega_2 = \omega_0 + \Delta_0$, where $\Delta_0 \ll \omega_0$.
In this case we can perform a rotating-wave approximation to single out the slowly-oscillating
coupling terms, and obtain the desired effective three-mode system, neglecting any higher-order non-linearities.
In the original frame, the Hamiltonian with this rotating-wave approximation is
\begin{eqnarray}\label{eq:interaction}
H &=& \sum_{k=0}^2 \omega_k a_k^\dagger a_k + i\kappa(a_1^\dagger a_2^\dagger a_0 - a_1a_2a_0^\dagger),
\end{eqnarray}
where $a_1$ and $a_2$ are the signal and idler modes, respectively, and $a_0$ is the pump mode.
Furthermore, we apply a driving force that is nearly resonant with $\omega_0$, with frequency $\omega_L = \omega_0 - \Delta_L$,
$|\Delta_L| \ll \omega_0$,
and transform the above Hamiltonian to the rotating frame where the resonant drive terms are time-independent,
\begin{eqnarray}
\label{eq:hamiltonian-3-mode}
H &=& \Delta_L a_0^\dagger a_0 + \sum_{k=1,2}\Delta_k a_k^\dagger a_k \nonumber\\
&+&
i\kappa(a_1^\dagger a_2^\dagger a_0 - a_1a_2a_0^\dagger)
 - i(Ea_0^\dagger - E^*a_0).
\end{eqnarray}
Here $\Delta_1 = \Delta_2 = (\Delta_0 - \Delta_L)/2$, $\kappa = \beta_{012}$ is the
inter-mode interaction strength, and $E$ is the driving amplitude of mode $a_0$. See Fig.~\ref{fig:modematching} for a
visual representation of the mode-matching condition and the detuning parameters $\Delta_0$ and $\Delta_L$.

This is an all-mechanical realization of the general three-mode parametric oscillator model in
nonlinear optics \cite{kheruntsyan:2000, mcneil:1983}, where mode $a_0$
is the quantized pump mode, and modes $a_1$ and $a_2$ are the signal and idler modes,
respectively.

\subsection{Effective two-mode model}\label{sec:paramp-regime-2mode}

We assume that in this purely nanomechanical realization of the parametric oscillator model, Eq.~(\ref{eq:hamiltonian-3-mode}),
all three mechanical modes interact with independent environments. We describe these processes with a standard Lindblad master equation on the form
\begin{eqnarray}
\dot\rho = -i[H, \rho] + \sum_k \gamma_k\left\{(N_k + 1) \mathcal{D}[a_k] + N_k \mathcal{D}[a_k^\dagger]\right\}\rho
\end{eqnarray}
where $\mathcal{D}[a_k]\rho = a_k \rho a_k^\dagger - \frac{1}{2} a_k^\dagger a_k \rho - \frac{1}{2} \rho a_k^\dagger a_k$
is the dissipator of mode $a_k$, $\gamma_k$ is the corresponding dissipation rate,
and the average thermal occupation number is $N_k =\left(\exp(\hbar\omega_k\beta) - 1\right)^{-1}$.
Here $\beta = 1/k_BT$ is the inverse temperature $T$, and $k_B$ is Boltzmann's constant.

\begin{figure}[t]
\begin{overpic}[width=\columnwidth]{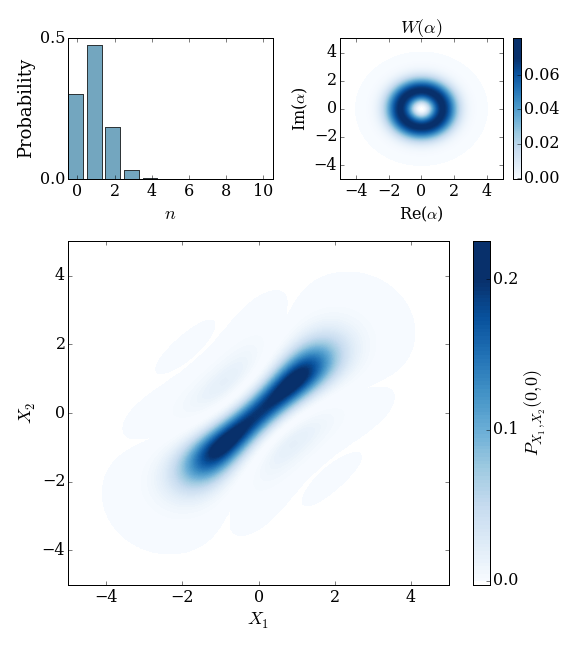}
\put(12,96){(a)}
\put(54,96){(b)}
\put(12,65){(c)}
\end{overpic}
\caption{
\label{fig:steadystate-visualization}
(Color online)
Visualization of the steady state of the effective two-mode system.
(a) The Fock state distribution of mode $a_1$ and $a_2$.
(b) The single-mode Wigner function of mode $a_1$ and $a_2$.
Both the Fock state distribution and the Wigner function are identical for
both modes $a_1$ and $a_2$, due to the symmetric two-phonon processes and
equal dissipation rates and initial states. The single-mode Wigner function
is positive, and the single-mode state can therefore be considered classical.
However, strong correlations exists between quadratures of two different modes,
as shown in the joint quadrature probability distribution $P_{X_1,X_2}(0,0)$ in
(c). Here we have used the parameters $\kappa = 0.15$, $E = 0.094$, $\gamma_0 = 1.0$,
$\gamma_1 = \gamma_2 = 0$, and $\Delta_0 = \Delta_L = 0$.
These parameters were chosen to produce a steady state
that closely corresponds to the ideal state\cite{munro:1999} for quadrature Bell inequality violation,
i.e. with $r=1.12$ (see Sec. IV.A).
}
\end{figure}

Assuming that the pump mode is strongly damped compared to the signal and idler
modes, $\gamma_0 \gg \gamma_1, \gamma_2$, and that the pump-mode dissipation dominates
over the coherent interaction, $\gamma_0 \gg \langle H \rangle \sim \kappa\langle a_0^\dagger a_1a_2 \rangle$,
one can adiabatically eliminate\cite{mcneil:1983,reid:1993,wiseman:1993} the
pump mode from the master equation given above.
Here we also assume that the high-frequency pump mode is at zero temperature, $N_0 = 0$, while the
temperatures of modes $a_1$ and $a_2$ can remain finite.
This results in a two-mode master equation that includes correlated two-phonon dissipation,
where one phonon from each mode dissipates to the environment through the pump mode,
in addition to the original single-phonon losses in each mode:
\begin{eqnarray}
\label{eq:me-eff-2-mode}
\dot\rho = &-&i[H', \rho] + \gamma \mathcal{D}[a_1a_2]\rho\nonumber\\
&+&
\sum_{k=1,2} \gamma_k \left\{(N_k + 1) \mathcal{D}[a_k] + N_k \mathcal{D}[a_k^\dagger]\right\}\rho,
\end{eqnarray}
where the effective two-phonon dissipation rate is
\begin{equation}
\gamma = \frac{\kappa^2\gamma_0/2}{|\gamma_0/2 + i\Delta_L|^2}.
\end{equation}
The reduced two-mode Hamiltonian is given by
\begin{eqnarray}
\label{eq:hamiltonian-eff-2-mode}
H' &=&
\sum_{k=1,2}\Delta_k a_k^\dagger a_k +
i(\mu a_1^\dagger a_2^\dagger - \mu^*a_1a_2) + \chi a_1^\dagger a_1 a_2^\dagger a_2,\nonumber\\
\end{eqnarray}
with the two-mode interaction strength
\begin{equation}
\mu = \frac{E\kappa}{\gamma_0/2 + i\Delta_L},
\end{equation}
and the effective cross-Kerr interaction strength
\begin{equation}
\chi = - \frac{\kappa^2\Delta_L}{|\gamma_0/2 + i\Delta_L|^2},
\end{equation}
which vanishes when the driving field is at exact resonance with the pump mode.
In the following we will generally assume that this resonance condition can
be reached, and $\Delta_L$ will be set to zero in the equations above.

In this resonant limit the Hamiltonian $H'$ describes an ideal two-mode parametric
amplifier, which is well-known to be the generator of two-mode squeezed states \cite{reid:1988}.
When applied to the vacuum state, or a low-temperature thermal state, the resulting two-modes squeezed states are nonclassically correlated \cite{marian:2003},
but when viewed individually, both modes appear to be in thermal states.
In spite of being quantum mechanically entangled, these two-mode
squeezed states have a positive Wigner function and cannot violate the
quadrature binning Bell inequalities \cite{gilchrist:1998} that we consider below.  One must consider
the effect of the two-phonon dissipation in Eq. [\ref{eq:me-eff-2-mode}] to induce such violations.

\begin{figure}[t]
\begin{overpic}[width=\columnwidth]{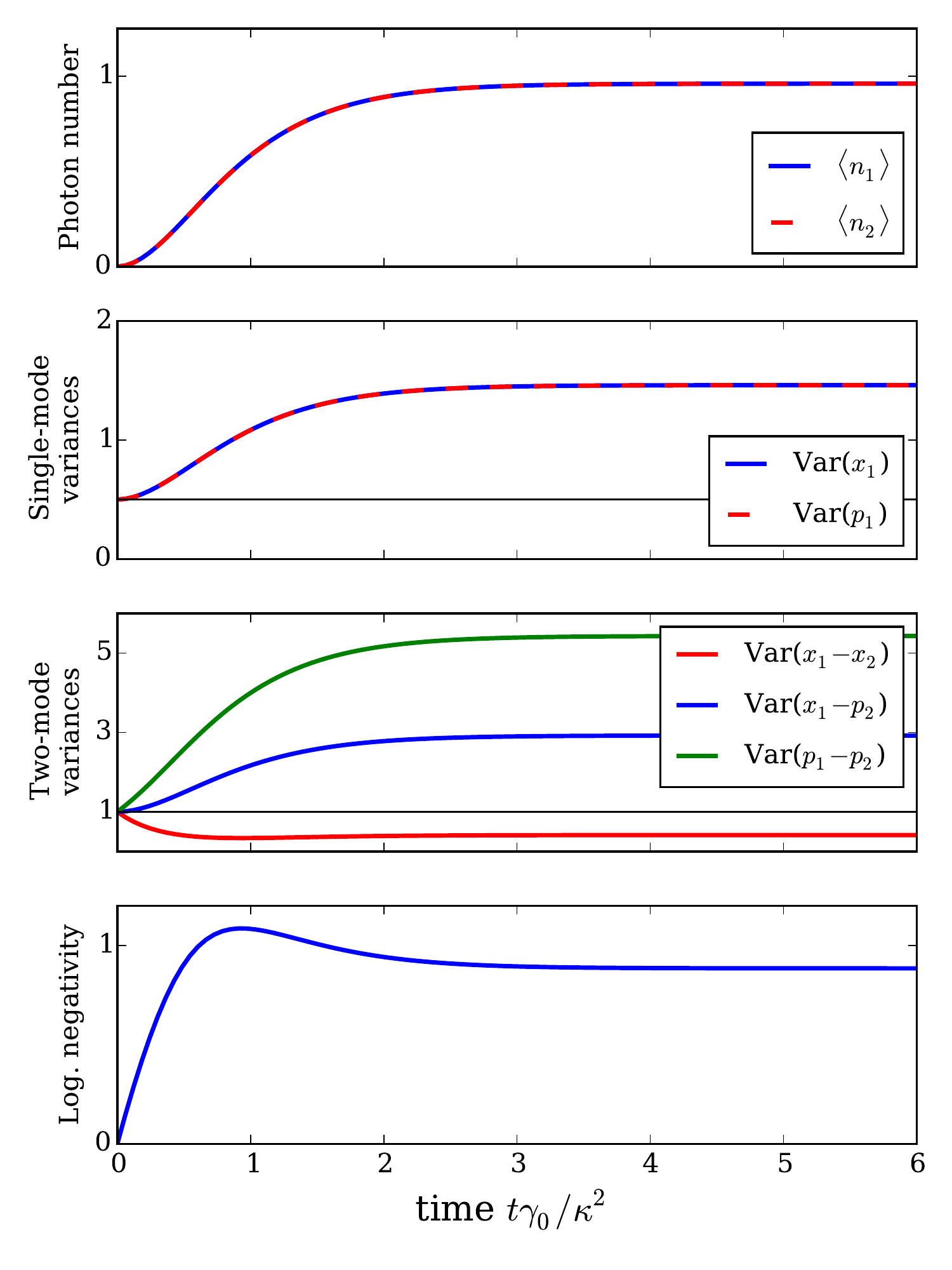}
\put(10.5,94){(a)}
\put(10.5,71){(b)}
\put(10.5,48){(c)}
\put(10.5,25){(d)}
\end{overpic}
\caption{
\label{fig:squeezing-vs-time}
(Color online)
The time evolution of the phonon number (a), single-mode quadrature variances (b),
and the variances of the two-mode quadrature differences (c), the
logarithmic negativity (d), for the case when the state is initially in the
zero-temperature ground state. In the large-time limit
the state approaches the steady state that is visualized in
Fig.~\ref{fig:steadystate-visualization}. The single-mode variances increase above
the vacuum limit as time increases, but the variance of cross-quadrature difference
Var$(x_1-x_2)$ decrease below the vacuum limit, which is a characteristic of two-mode
squeezing, and the nonzero logarithmic negativity demonstrates that the steadystate
is nonclassical. Here we used the same parameters as those given in the caption
of Fig.~\ref{fig:steadystate-visualization}.
}
\end{figure}
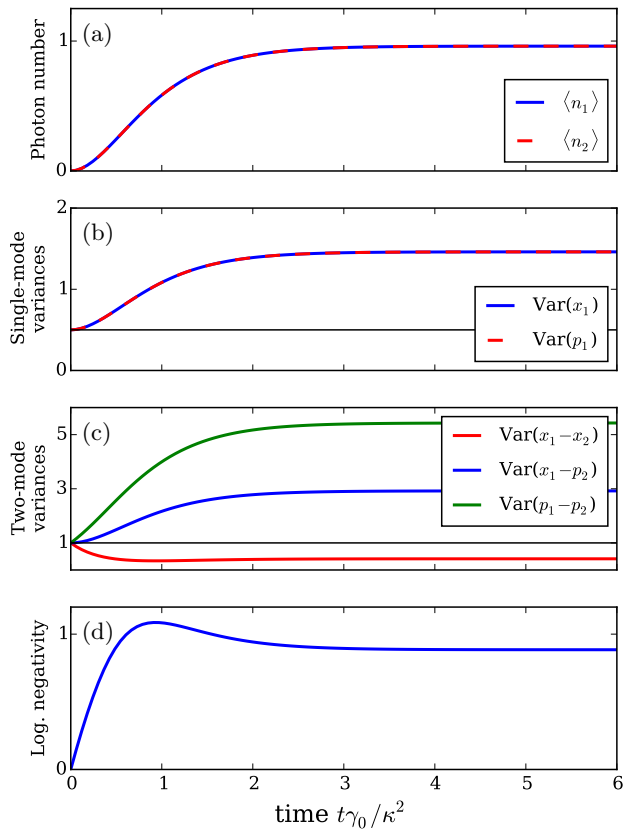

In the highly idealized case when single phonon dissipation in the $a_1$ and $a_2$ modes is absent,
i.e., $\gamma_1 = \gamma_2 = 0$, but with $\gamma_0>0$, the model Eq.~(\ref{eq:me-eff-2-mode})
produces a steady state \cite{reid:1993} of the form
\begin{equation}
 \label{eq:ideal_steady_state}
 \rho = \frac{1}{I_0(2r^2)}\sum_{m,n}\frac{r^{2m+2n}}{m!n!}|m,m\rangle\langle n,n|,	
\end{equation}
where $I_0$ is the zeroth order modified Bessel function and $r = \sqrt{2E/\kappa}$. The special structure of this steady state,
with equal number of phonons in each mode, is because both the
Hamiltonian and two-phonon dissipator conserve the phonon-number difference
$a_2^\dagger a_2 - a_1^\dagger a_1$. However, this symmetry is broken if the single-phonon
dissipation processes are included in the model, i.e. $\gamma_1, \gamma_2 > 0$.
The state Eq.~(\ref{eq:ideal_steady_state}) is visualized in Fig.~\ref{fig:steadystate-visualization}, for the specific
set of parameters given in the figure caption. Figure \ref{fig:steadystate-visualization}(a-b) show the
Fock-state distribution and the Wigner function for the modes $a_1$ and $a_2$ (because of symmetry
the states of both modes are identical in this case, and only one is shown).
In this case the states of the two modes no longer appear to be thermal
when viewed individually, but the reduced single-mode Wigner functions are positive and thus, on their own, each mode appears classical. However, together, the two-mode Wigner function can be negative.  For example, there is a strong cross-quadrature correlation, as shown in
Fig.~\ref{fig:steadystate-visualization}(c). The variances of the cross-quadrature differences, in the transient
approach to the steady state, are shown in Fig.~\ref{fig:squeezing-vs-time}, and exactly in the steady state the variance of
the squeezed two-mode operator difference is
\begin{equation}
	{\rm Var}(x_1-x_2) = 1 + 2r^2\frac{I_1(2r^2)}{I_0(2r^2)} - 2r^2,
\end{equation}
which in the limit of large $r$ approaches 1/2, but has a local minimum of about 0.4 at $r \approx 0.92$. We note that
for the vacuum state ${\rm Var}(x_1-x_2) = 1$, and thus this quadrature difference variance is therefore
squeezed below the vacuum level for any $r>0$. The logarithmic negativity \cite{adesso:2007} shown in Fig.~\ref{fig:squeezing-vs-time}(d) further
demonstrates the nonclassical nature of this state.

These intermode quadrature correlations, with squeezing below the vacuum level of fluctuations, are nonclassical and it has been shown
that this particular state can violate Bell inequalities based on quadrature measurements \cite{gilchrist:1998}, as we will discuss in the next section.
In fact, this steady state is, for a certain value of $r$, a good approximation to the ideal two-mode quantum state \cite{munro:1999} for these
kind of Bell inequalities. However, it has also been shown that in the presence of single phonon dissipation
the steady state two-mode Wigner function is always positive, and
thus exhibits a hidden-variables description and cannot violate any Bell inequalities
\cite{kheruntsyan:2000}. Fortunately, this is only the case for the steady state, and there can be a
significant transient period in which the two-mode system is in a state that can give a violation.

In the following we consider two regimes; the steady state, and the slow transient dynamics of a system that is originally
in the ground state, and approaches the new steady state after the driving field has been turned on.

\section{Bell inequalities for nanomechanical systems}\label{sec:quadrature-bell}

Verifying that a nanomechanical system is in the quantum regime, and that the states produced in the system are nonclassical,
can be sometimes be experimentally challenging, largely because of the difficulty in implementing single-phonon detectors in nanomechanical systems.
As has been done in circuit QED \cite{fink:2008, schuster:2008}, measuring a nonlinear energy spectrum \cite{rips:2013} would be a convincing
indication that the system is operating in the quantum regime, although it does not imply that the state of the system
is nonclassical, and all quantum nanomechanical systems need not necessarily be nonlinear.
A number of techniques could be used to demonstrate that the state is nonclassical \cite{miranowicz:2010}, for example reconstructing
the Wigner function using state tomography and looking for negative values,
or evaluating entanglement measures such as the logarithmic negativity (for Gaussian states) or
entanglement entropy (suitable only at zero temperature).

Here we are interested in a nonclassicality test that can be evaluated using joint two-mode quadrature measurements.  The two-mode squeezing
shown in the previous section can be considered as an entanglement witness\cite{simon:2000,duan:2000}, and was recently investigated experimentally in an opto-mechanical device\cite{palomaki:2013,palomaki:2013:2}.
The quadrature-based Bell inequality can be seen as another, stricter, example of a nonclassicality test, and in the following we focus on the possibility
of violating these Bell inequalities with the nanomechanical system outlined in the previous section.
Even though one cannot rule out the locality-loophole in such a system, and thus a violation would lack any meaning as a strict test of Bell nonlocality \cite{brunner:2013}, it would still serve as a very satisfying test for two-mode entanglement.

The original Bell inequalities are formulated for dichotomic measurements, with two possible outcomes.
However, dichotomic measurements are not normally available in harmonic systems like the nanomechanical systems considered here,
where the measurement outcomes are, for the most part, continuous and unbound.
In this continuous-variable limit one must choose how to perform a Bell inequality test with care. Generalized inequalities for unbound measurements exist\cite{cavalcanti:2007}, but are both extremely challenging to implement and hard to violate. Fortunately one can implement CHSH-type Bell inequality by binning quadrature measurements, and thus obtaining a dichotomic bound observable. Munro \cite{munro:1999} showed that, while in general it is hard to generate states which can violate such an inequality, it is possible to generate precisely the type of states which do cause a violation with a nondegenerate parametric oscillator, which is analogous to the system we investigate here.

One possible binning strategy\cite{gilchrist:1998,wenger:2003} for the continuous outcomes of
quadrature measurements of the mechanical modes is to classify the outcomes as $1$
if the measurement outcome is $X_\theta > 0$, and $0$ otherwise. The probability of the
outcomes $0$ and $1$ for the two modes can then be written
\begin{eqnarray}
\label{eq:p_alpha_beta}
P_{\alpha\beta}(\theta, \phi) &=& \int_{L(\alpha)}^{U(\alpha)}\int_{L(\beta)}^{U(\beta)} d^2X p(X_1^\theta, X_2^\phi)[\rho]
\end{eqnarray}
where
\begin{equation}
L(\alpha) = \begin{cases}
0 & \text{if $\alpha = 1$} \\
-\infty & \text{if $\alpha = 0$}
\end{cases},
U(\alpha) = \begin{cases}
\infty & \text{if $\alpha = 1$} \\
0 & \text{if $\alpha = 0$}
\end{cases}.
\end{equation}
Here $\rho$ is the two-mode density matrix and $p(X_1^\theta, X_2^\phi)[\rho]$ is the
probability distribution for obtaining the measurement outcomes $X_1^\theta$ and $X_2^\phi$
for the signal and idler mode quadratures
\begin{eqnarray}
 x_1^\theta &=& a_1 e^{-i\theta} + a_1^\dagger e^{i\theta}, \\
 x_2^\phi &=& a_2 e^{-i\phi} + a_2^\dagger e^{i\phi},
\end{eqnarray}
respectively.
This probability distribution is given by
\begin{eqnarray}
\label{eq:p_x1_x2_rho}
&&p(X_1^\theta, X_2^\phi)[\rho]
= \left<X_1^\theta, X_2^\phi\right|\rho\left|X_1^\theta, X_2^\phi\right> \nonumber\\
&=&
\sum_{m,n,p,q}\rho_{(m,n),(p,q)}
\frac{e^{-i(m\theta+n\phi)}e^{i(p\theta+q\phi)}}{\pi\sqrt{2^{m+n+p+q} m!n!p!q!}} \nonumber\\
&& \times e^{-X_1^2}e^{-X_2^2}H_m(X_1)H_n(X_2)H_p(X_1)H_q(X_2) \nonumber\\
\end{eqnarray}
where $H_n(x)$ is the Hermite polynomial of $n$th order, and where we have written
the density matrix in the two-mode Fock basis,
\begin{eqnarray}
\rho =
\sum_{m,n,p,q}\rho_{(m,n),(p,q)} \left|m,n\rangle\langle p,q\right|.
\end{eqnarray}
The integral in Eq.~(\ref{eq:p_alpha_beta}) can be evaluated analytically\cite{munro:1999},
but in general the sum in Eq.~(\ref{eq:p_x1_x2_rho}) cannot.

Treating the binned quadrature measurements as dichotomic observables we can write the
standard Bell's inequalities in the Clauser-Horne (CH) \cite{clauser:1974} form
\begin{equation}
\label{eq:bell_ch}
B_{\rm CH} = \frac{P_{11}(\theta, \phi) - P_{11}(\theta, \phi') + P_{11}(\theta', \phi) + P_{11}(\theta', \phi')}{P_{1}(\theta') + P_{1}(\phi)}
\end{equation}
which for a classical state satisfies $|B_{\rm CH}| \leq 1$,
and in the Clauser-Horne-Shimony-Holt (CHSH) \cite{clauser:1969} form
\begin{eqnarray}
\label{eq:bell_chsh}
B_{\rm CHSH} &=& E(\theta, \phi) - E(\theta', \phi) + E(\theta, \phi') + E(\theta', \phi'),\;\; \\
E(\theta, \phi) &=& P_{11}(\theta, \phi) + P_{00}(\theta, \phi) \nonumber\\
&-& P_{10}(\theta, \phi) - P_{01}(\theta, \phi),
\end{eqnarray}
which for a classical state satisfies $|B_{\rm CHSH}| \leq 2$. Here we have also used
\begin{eqnarray}
P_{1}(\theta) &=& \int_0^\infty\int_{-\infty}^\infty d^2X p(X_1^\theta, X_2^\phi)[\rho].
\end{eqnarray}

Both $B_{\rm CH}$ and $B_{\rm CHSH}$ are in general functions of the four angles $\theta, \phi, \theta'$, and $\phi'$.
However, to reduce the number of parameters here we consider the angle parameterization $\theta = -2\varphi$, $\phi=3\varphi$,
$\theta'=0$, and $\phi'=\varphi$, which only leaves a single free angle parameter $\varphi$. In principle this can reduce
the magnitude of violation one can observe, but as we will see this parameterization still allows violations to occur for
the types of states we are interested in here. In the following we evaluate both $B_{\rm CH}$ and $B_{\rm CHSH}$
using this angle parameterization.

\section{Violation of Bell's inequality with nanomechanical resonators}\label{sec:mechanical-bell-violation}

In this section we investigate the conditions under which the states formed in the multimode nanomechanical system
may violate Bell's inequality. We emphasize again that in this context we are interested
in Bell's inequality as a test that can demonstrate
entanglement between different mechanical modes. We begin with an analysis of the
steadystate for the idealized model with $\gamma_1,\gamma_2 = 0$,
and then turn our attention to the transient behaviour for finite $\gamma_1$ and
$\gamma_2$.

\subsection{Steady state}

With $\gamma_1,\gamma_2 = 0$, the steady state is given by Eq.~(\ref{eq:ideal_steady_state}),
and inserting this state in the Bell inequalities Eqs.~(\ref{eq:bell_ch}-\ref{eq:bell_chsh})
gives an expression as a function of the steady state parameter $r$ and the angle $\varphi$
that can be optimized for maximum Bell violation. The optimal value of the angle turns out to
be $\varphi = \pi/4$, and the resulting equation for optimal $r$ is
\begin{equation}
\label{eq:optimal_r}
I_0(2r^2) \frac{dG(r)}{dr} = 4r^2I_1(2r^2) G(r),
\end{equation}
but the sum over Fock-state basis that comes from Eq.~(\ref{eq:p_x1_x2_rho})
cannot to our knowledge be evaluated in a simple analytical form, so we have
\begin{eqnarray}
G(r) &=&
\sum_n\sum_{m>n} \frac{8(2r^2)^{n+m}\pi}{(n!m!)^2(n-m)^2}
\left[\mathcal{F}(n,m) - \mathcal{F}(m,n)\right]^2 \times \nonumber\\
&&\left\{3\cos\left[(n-m)\varphi\right] -  \cos\left[3\varphi(n-m)\right]\right\},
\end{eqnarray}
and
\begin{equation}
\mathcal{F}(n,m) = \left[\Gamma\left(\frac{1}{2}-\frac{n}{2}\right)\Gamma\left(-\frac{m}{2}\right)\right]^{-1},
\end{equation}
as given in Ref.~\onlinecite{munro:1999}. Solving Eq.~(\ref{eq:optimal_r}) numerically
gives $r_{\rm opt} \approx 1.12$, as reported in Ref.~\onlinecite{gilchrist:1998}.
The corresponding steady state Eq.~(\ref{eq:ideal_steady_state}) for $r_{\rm opt}$ is visualized
in Fig.~\ref{fig:steadystate-visualization}. We note that for this optimal Bell violating state
the mean phonon number in each mode is only $\langle n \rangle \approx 0.94$, which highlights
the need to operate the system near its ground state. If fact, when $\langle n \rangle \gg 1$ no
Bell inequality violation can be observed.

In our nanomechnical model this translates to an optimal driving strength,
$E_{\rm opt} = r_{\rm opt}^2\kappa/2$,
that maximizes the Bell inequality violation for a given nonlinearity $\kappa$.
This optimal driving amplitude $E_{\rm opt}$ applies to the steady state of the
idealized model without single-phonon dissipation. With finite single-phonon
dissipation, the steady state does not violate any of the Bell inequalities.
However, as we will see in the following section, $E_{\rm opt}$ still gives a good approximation
for the optimal transient violation.  While these transients are harder to capture, recent experiments on
opto-mechanical systems have shown they are in principle possible \cite{palomaki:2013,palomaki:2013:2}, and relevant for the alternative proposal in the final section below.  How
far one can go with using multiple ancilla optical or microwave cavities to perform similar measurements on different internal modes of a single mechanical device is not yet clear.

\subsection{Transient}
\begin{figure}[t]
\begin{overpic}[width=\columnwidth]{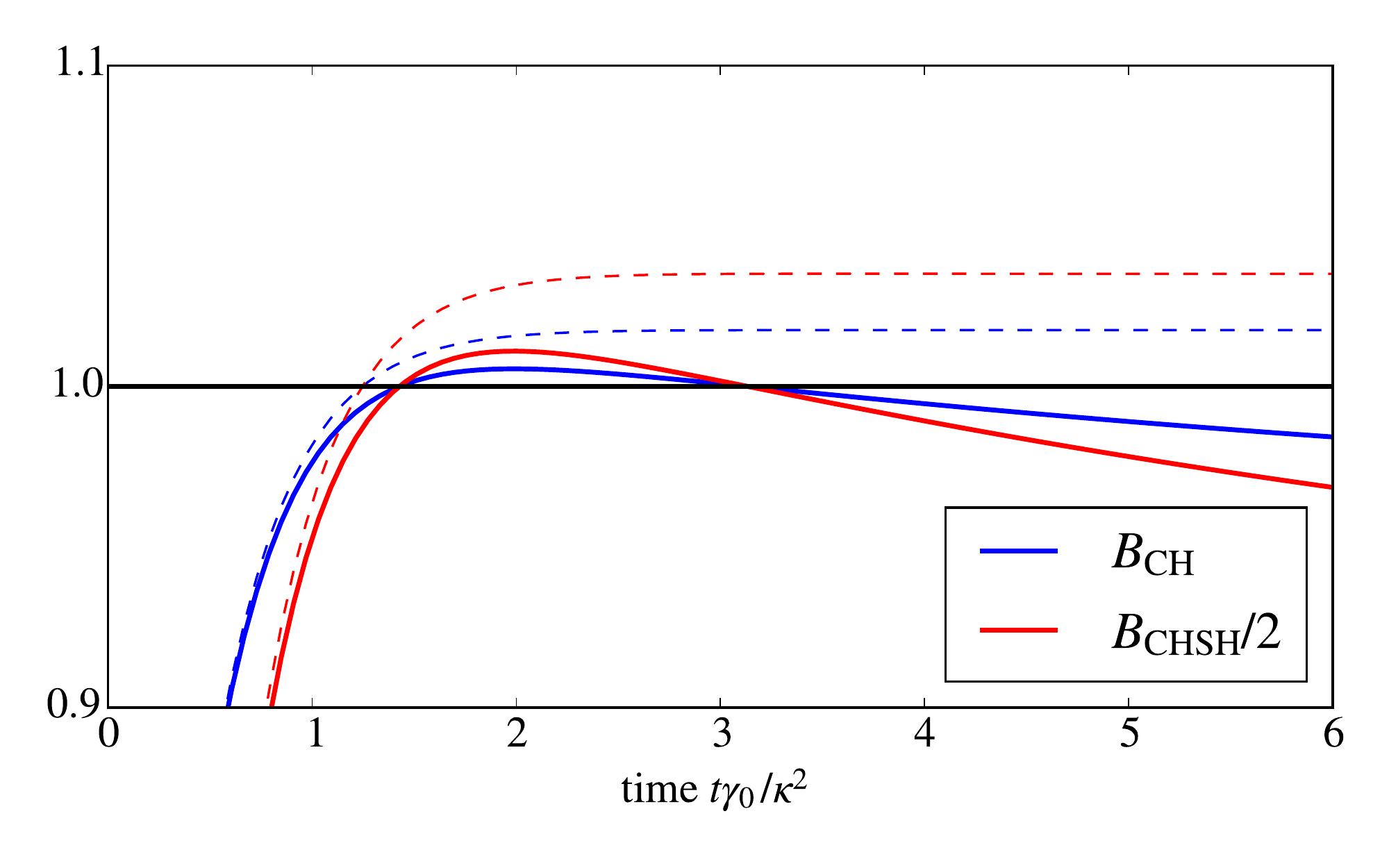}\put(10,62){(a)}\end{overpic}
\begin{overpic}[width=\columnwidth]{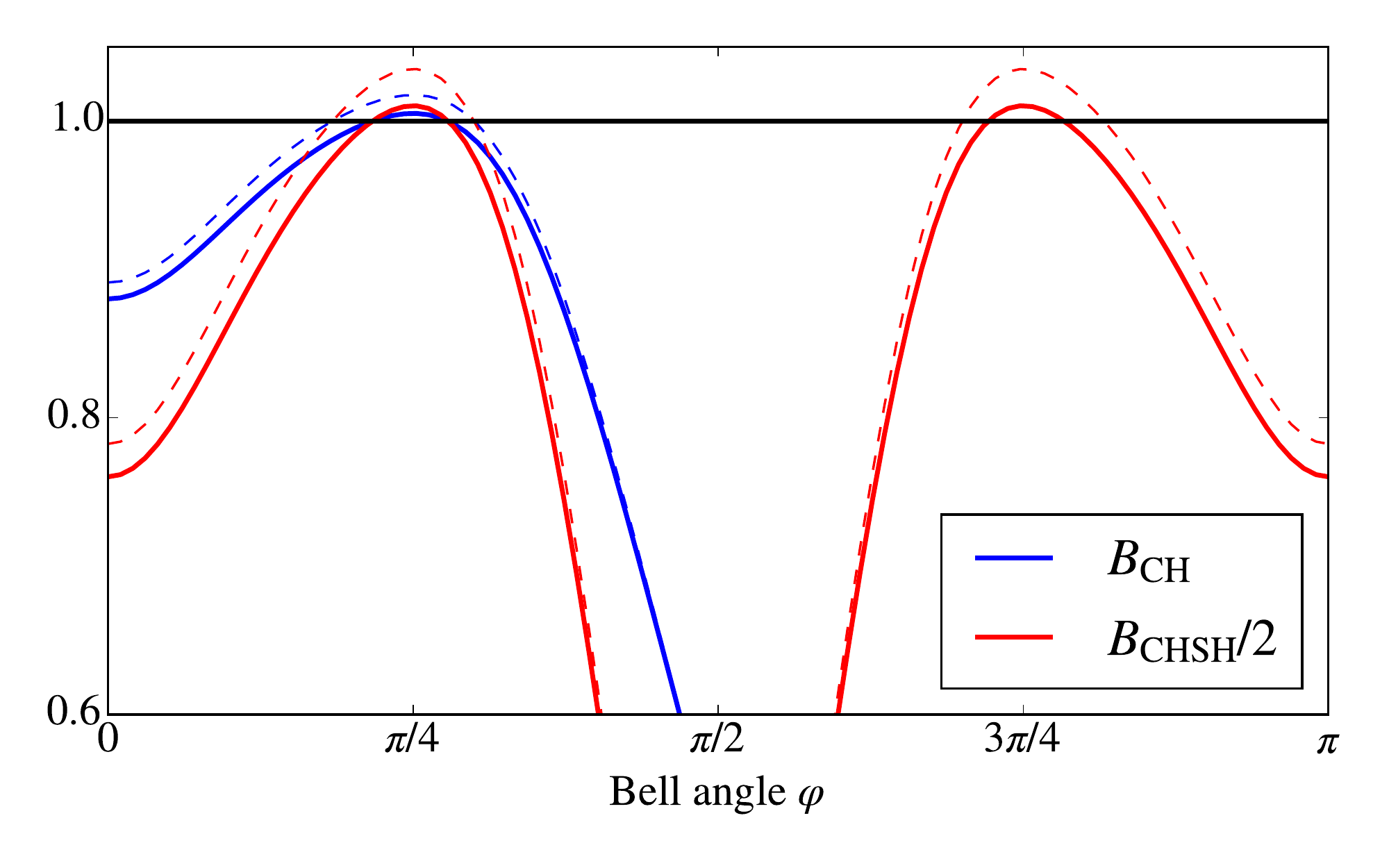}\put(10,64){(b)}\end{overpic}
\caption{
\label{fig:bell-violation-vs-time}
(Color online) (a) The normalized CH and CHSH Bell quantities (violation
above 1) as a function of time, for the pumped two-mode nanomechanical
resonator, under the ideal condition without signal and idler mode dissipation (dashed)
and for the case including signal and idler dissipation.
The initial state is the vacuum state, which we assume can be prepared to a good
approximation using cooling. At $t=0$, the parametric
amplification is turned on by the activation of the driving field with amplitude $E$.
In the ideal case, the steady state violates both the CH and CHSH Bell inequalities,
but there is no steady state violation when single-phonon dissipation is included. However,
there is a period of time during the transient where both inequalities are violated.
(b) The angle $\varphi$ dependence for the normalized CH and CHSH Bell quantities for
$t\gamma_0/\kappa^2 = 1.8$, where solid lines include dissipation and dashed lines are the ideal case.
The optimal value of $\varphi$ for the states produced in the model we investigate here
is $\varphi=\pi/4$, which was used in (a). The parameters used here are the same as in
Fig.~\ref{fig:steadystate-visualization}, and for the solid lines we used $\gamma_1=\gamma_2=0.001$.
}
\end{figure}

\begin{figure*}[t]
\begin{overpic}[width=0.64\columnwidth]{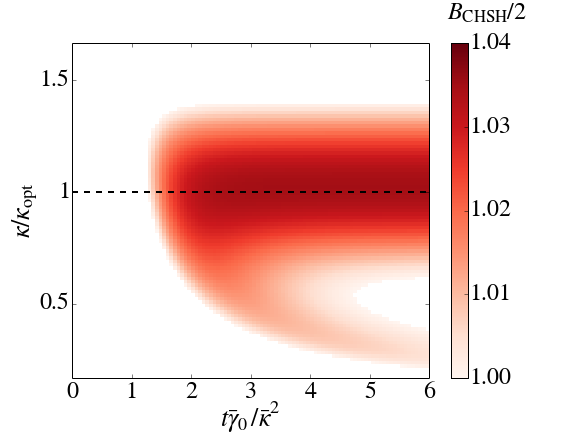} \put(-3,60){(a)}\end{overpic}
\begin{overpic}[width=0.64\columnwidth]{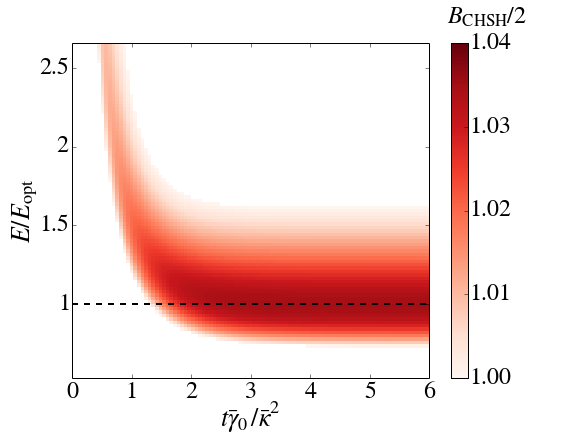} \put(-3,60){(b)}\end{overpic}
\begin{overpic}[width=0.64\columnwidth]{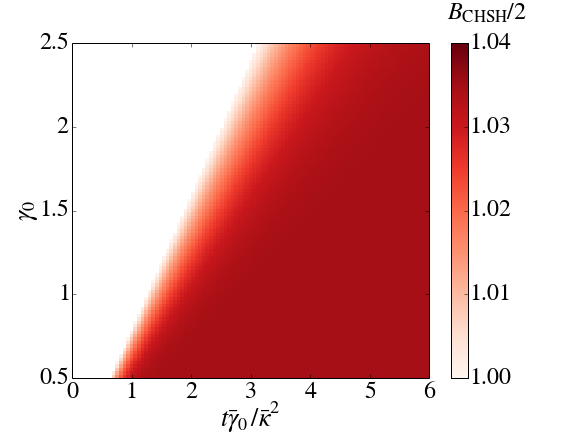} \put(-3,60){(c)}\end{overpic}
\begin{overpic}[width=0.64\columnwidth]{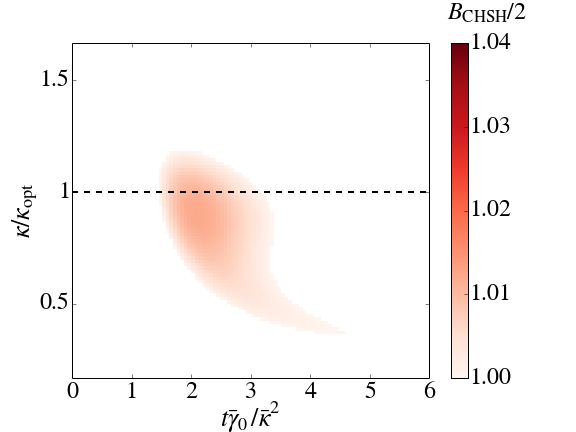} \put(-3,60){(d)}\end{overpic}
\begin{overpic}[width=0.64\columnwidth]{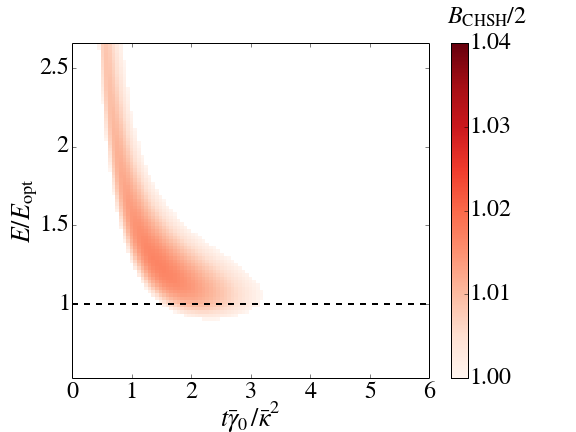} \put(-3,60){(e)}\end{overpic}
\begin{overpic}[width=0.64\columnwidth]{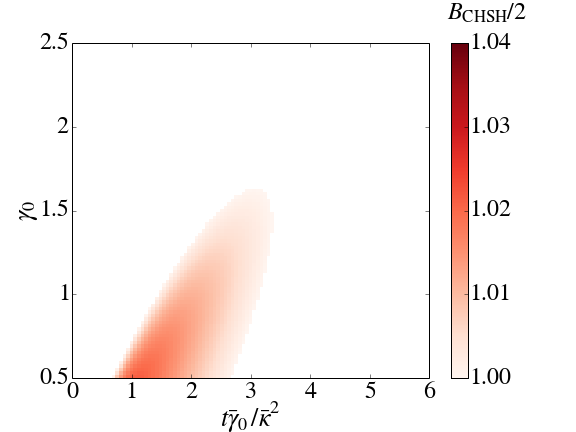} \put(-3,60){(f)}\end{overpic}
\caption{
\label{fig:bell-violation-transient}
(Color online)
Violation of the normalized quadrature CHSH Bell inequality (redish region) as a function of
time $t$ and the inter-mode coupling $\kappa$ (a,d), the pump mode driving amplitude $E$ (b,e),
and the pump-mode dissipation rate $\gamma_0$ (c,f).
The ideal case without signal and idler mode dissipation is shown in (a-c), and
(d-e) include signal and idler mode dissipation with equal dissipation rates
$\gamma_1=\gamma_2 = 0.001$.
In (a-c) there is a parameter window for $\kappa$ and $E$ which results in a violation
for sufficiently large $t$, as well as in the steady state. However, in (d-e) there is
no violation in the steady state, but during a transient time a violation may still occur
for suitably chosen parameters.
Apart from the parameters on the vertical axes, the parameters were kept fixed at the same
values as given in Fig.~\ref{fig:steadystate-visualization}, and
denoted by a bar over the symbol in the axes.
}
\end{figure*}

Since the more realistic model, with finite single-phonon dissipation processes, does not produce
a steady state that violates any of the Bell inequalities, we are lead to investigate
transient dynamics. Here we focus on the transient which occurs when the driving field $E$ is
turned on after the relevant modes have been cooled to their ground states. The state of the
system then evolves from the ground state to the steady state that does not violate the Bell
inequalities. However, if the single phonon dissipation processes are sufficiently slow there can be
a significant time interval during which the state of the system does violate the Bell inequalities.

To investigate this transient dynamics we numerically evolve the effective two-mode system described
by the master equation, Eq.~(\ref{eq:me-eff-2-mode}), and
evaluate the $B_{\rm CH}$ and $B_{\rm CHSH}$ quantities as a function of time and the angle $\varphi$.
The results shown in Fig.~\ref{fig:bell-violation-vs-time},
for the situations with and without signal and idler mode dissipation and at zero temperature,
demonstrate that the nanomechanical system we consider can indeed be driven into a transient state that
violates both types of Bell inequalities. With losses the onset of violation
is proportional to $\gamma_0/\kappa^2$, and the time at which the violation cease is proportional
to $\gamma_1^{-1}, \gamma_2^{-1}$, so if
\begin{equation}
\label{eq:transient_violation_regime}
\gamma_1,\gamma_2 \ll \kappa^2/\gamma_0,
\end{equation}
we expect a significant period of time during the transient where the inequalities will be violated.
We note that in Fig.~\ref{fig:bell-violation-vs-time}(a), the regions of violation for the CH and CHSH
inequalities are identical, and this is, according to our observations, always the case for this model and angle parametrization.
Because of this, in the following we only show the results for the CHSH inequality.

To further explore the parameter space that can produce a Bell-inequality violation we
evolve the master equation as a function of time and the parameters $E$, $\kappa$ and $\gamma_0$, for both the
ideal case with dissipation-less signal and idler modes, $\gamma_1=\gamma_2=0$, and for the case including
signal and idler mode dissipation, $\gamma_1,\gamma_2 > 0$. In these simulations the initial state is always the
ground state, and we take the temperature of the signal and idler modes to be zero.
The results are shown in Fig.~\ref{fig:bell-violation-transient}(a-c) and (e-f), respectively.
From Fig.~\ref{fig:bell-violation-transient} it is clear that for the case $\gamma_1=\gamma_2=0$,
there exist optimal values of $\kappa$ and $E$, given that other parameters are fixed,
that produce steady states that maximally violates the Bell inequality (marked with dashed lines in the figures).
However, importantly, we also note that the optimal values for $\kappa$ and $E$ for the steady state of
the ideal model also give a good indicator for the optimal regime for the Bell violation in the transient of the
case with finite single-phonon dissipation, when additionally taking into account the time
scales for the transient given in Eq.~(\ref{eq:transient_violation_regime}).

\begin{figure}[t]
\includegraphics[width=\columnwidth]{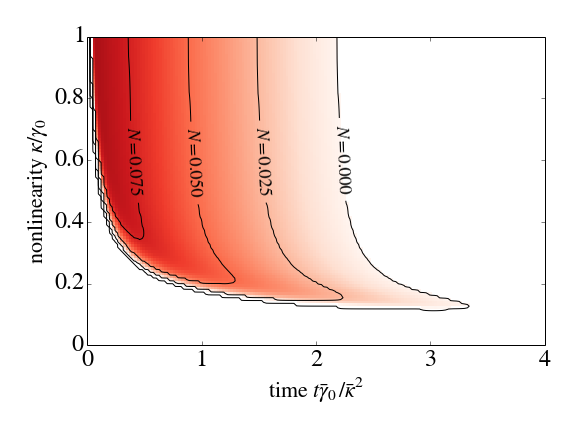}
\caption{
\label{fig:bell-violation-transient-vs-temperature}
(Color online) The regions where a transient Bell inequality violation can be achieved,
as a function of nonlinearity $\kappa$ and time $t$, and for different temperatures. The region of violation
at zero temperature is shown in redish color. The contours mark the regions of violation for finite signal and idler
mode temperatures (assumed equal), labeled by the average number $N$ of thermal phonons.
We note that even a very small number of initial thermal phonons inhibits the Bell inequality violation,
which suggest that excellent ground-state cooling of both modes is a prerequisite to obtaining a violation.
}
\end{figure}

When the signal and idler modes have finite temperature the region of Bell inequality violation is further reduced, as
shown in Fig.~\ref{fig:bell-violation-transient-vs-temperature}. The detrimental effects of thermal phonons are two-fold:
It reduces the transient time-interval during which a violation can be observed, and the nonlinear
interaction strength required to be able to see any violation at all increases. In fact, to observe
a transient Bell inequality violation, the average thermal occupation number must be very small: An average thermal
occupation number of even $0.1$ phonon in the signal and idler mode is sufficient to inhibit any Bell violation with the
system we have considered here. Excellent ground state cooling is therefore a prerequisite to violating a Bell
inequality tests in a nanomechanical resonator.

\section{Experimental outlook}\label{sec:expr-outlook}

As can be seen in Fig.~\ref{fig:bell-violation-transient} and Fig.~\ref{fig:bell-violation-transient-vs-temperature},
the violation of a Bell inequality in the system we consider here requires, as expected, a combination of low temperature, large nonlinearity,
and transient quadrature measurements. These conditions can all be rather challenging to satisfy in an
experimental system, but on the other hand they are exactly the type of conditions that one can expect would have
to be satisfied for realistic quantum mechanical applications in these devices. The Bell inequality violation
can therefore be seen as a benchmark that indicates that entangled quantum states can be generated and detected
with high precision.

While one can imagine cryogenics and side-band cooling techniques can satisfy the first criteria \cite{oconnell:2010, teufel:2011:2, chan:2011,grajcar:2008},
the ultimate upper limit of the strength of intrinsic nonlinearities in mechanical systems is not clear.
In a recent experiment extremely large nonlinear intra-mode coupling was observed in a
carbon nanotube system when the modes had frequencies which were integer multiples of each other \cite{sampaz:2003}.
In that case a strong effective mode-mode coupling was also found, which is required for generating the Bell inequality violating states we consider here.
Nonlinear mode coupling has also been demonstrated and analyzed in doubly-clamped beam resonators \cite{westra:2010,lulla:2012,khan:2013,yamaguchi:2013}, and
circular graphene membrane resonators \cite{eriksson:2013}.
Also, recent studies\cite{rips:2013} have proposed enhancing the nonlinearity per phonon by reducing the fundamental
frequency of the mechanical oscillator, which essentially amounts to increasing the ground state displacement.
Thus, there is progress in realizing nonlinear mode interaction in several types of nanomechanical systems,
and sufficiently strong nonlinearities to produce Bell inequality violating states should be obtainable in these devices,
although further progress in this experimental work in this direction may be required.

Performing transient quadrature measurements of selected modes of the mechanical resonator is an another experimental
challenge. However, the displacement of a nanomechanical resonator can be converted to electrical signals and measured
for example using a range of different techniques, for example piezoelectric schemes \cite{mahboob:2012},
coupling to auxiliary optical modes \cite{bochmann:2013,paternostro:2007},
or by capacitive coupling to a microwave circuit \cite{regal:2008}. In recent experiments, transient quadrature measurements
of a nanomechanical system were carried out with high precision and level of control \cite{palomaki:2013,palomaki:2013:2}.
Aslo, in microwave electronics, quadrature measurements in the quantum regime have been applied to measure two-mode squeezing
\cite{eichler:2011:b,bergeal:2012}, state tomography \cite{mallet:2011}, and entanglement \cite{flurin:2012}. Given a
sufficiently efficient transducer from mechanical displacement to electrical signals, the outlook for the required
measurements for evaluating the quadrature Bell inequality is therefore encouraging.

\subsection{Optomechanical realization}\label{sec:expr-outlook2}

As an alternative to the purely mechanical scheme discussed so far one could observe the similar quadrature-based Bell inequality violations in an optomechanical setup akin to that proposed in
Refs.~\onlinecite{stannigel:2011} and \onlinecite{ludwig:2012}, where a single mechanical mode is coupled to two optical cavities, e.g., in a membrane-in-the-middle geometry or within a photonic crystal cavity. The most straightforward implementation
would be to use the mechanical mode as the pump mode which then acts to entangle the optical modes.
The optical modes are coupled due to a photon tunneling, and the resulting hybridized modes replace the mechanical signal and idler modes $a_1$ and
$a_2$ we discussed in this work. On resonance this again leads to the same
interaction we use in Eq. [\ref{eq:interaction}]. The main motivation of inducing this interaction in these earlier works
was to engineer anharmonic energy levels. This anharmonicity allows specific
transitions to be addressed with external laser fields allowing one to use such devices as single-phonon/photon transistors and for non-demolition
measurements of phonons or photons. In the limit where the mechanical pump mode can be driven and adiabatically eliminated, one in principle
observe Bell inequality
violations in the (hybridized) quadrature measurements of the two optical cavities.

\section{Combining even and odd nonlinearities: coupling mechanical qubits}\label{sec:mechanical-qubit}

In the previous calculations we have been exclusively considering the effect of odd nonlinearities which can only arise in  asymmetric mechanical systems.  In purely symmetric devices the even order terms dominate, arguably the most important of which is the $x^4$ Duffing nonlinearity. Recent works \cite{rips:2013} have examined how this induces an anharmonic energy spectrum in the fundamental mode of a nanomechanical system, and outlined how this anharmonic spectrum can be used as an effective qubit for quantum computation. Naturally one can consider the effect of both the third order coupling we have outlined here, and the third and fourth order Duffing self-anharmonicity. Ultimately the relative strengths of these different terms depend strongly on the overlap between the different mode shapes within the device, the geometry of the device, and the effect of various nonlinearity enhancing mechanisms.  A naive investigation of the contributions from these quartic terms suggest they only work to degrade the Bell inequality violation we discuss here.  However, going beyond the regime we have outlined thus far, one may note that, by changing the frequency of the driving field in Eq.~(\ref{eq:full_hamiltonian}) one can get an excitation-preserving beam-splitter type of interaction between the signal and idler modes.
\begin{equation}
H_{\rm int}^{c} = \mu (a_1^\dagger a_2+ a_2^\dagger a_1).
\end{equation}
If this is combined with a sufficiently strong third or fourth-order self nonlinearity, such that the lowest lying energy states of each mode can be considered as a two-level system, one has a means to couple different mechanical qubits in a single device. It may be possible to construct similar interactions with ancilla cavities and optomechanical interactions \cite{stannigel:2011,jacobs:2012}. The original parametric interaction described in Eq.~(\ref{eq:hamiltonian-3-mode}) is not useful for this purpose as it takes one out of a single excitation subspace, as does the two-phonon dissipation.

\section{Conclusions}\label{sec:conclusion}

We have investigated a regime of a multimode nanomechanical resonator, with intrinsic nonlinear mode coupling, in which three selected modes
realize a parametric oscillator. In the regime where the pump mode of the parametric oscillator can be adiabatically eliminated, we have
investigated the generation of entangled states between two distinct modes of oscillation in the nanomechanical resonator,
and the possibility of detecting this entanglement using quadrature-based Bell inequality tests. Our results demonstrate that while realistically
it will not be possible to violate any Bell inequality in the steady state, there can be a significant duration of time in which the transient evolution
from the ground state (prepared by cooling) to the steady state where the state of the system violates Bell inequalities.
However, to achieve this transient violation requires a relatively large nonlinear mode coupling, excellent ground state cooling, and fast and
efficient quadrature measurements. These are, of course, very challenging experimental requirements, but we believe that if a quadrature Bell inequality
violation is realized experimentally it would be a very strong demonstration of quantum entanglement in a macroscopic mechanical system.

\section*{Acknowledgements}

The numerical simulations were carried out using QuTiP \cite{johansson:2012, johansson:2013},
and the source code for the simulations are available in Ref.~[\onlinecite{figshare}]. 
We acknowledge W. Munro and T. Brandes for discussions and feedback.
This work was partly supported by the RIKEN iTHES Project, MURI Center for Dynamic Magneto-Optics, 
JSPS-RFBR No.~12-02-92100, Grant-in-Aid for Scientific Research (S),
MEXT Kakenhi on Quantum Cybernetics, the JSPS-FIRST program,
and JSPS KAKENHI Grant No.~23241046.

\bibliography{references}

\begin{thebibliography}{75}
\expandafter\ifx\csname natexlab\endcsname\relax\def\natexlab#1{#1}\fi
\expandafter\ifx\csname bibnamefont\endcsname\relax
  \def\bibnamefont#1{#1}\fi
\expandafter\ifx\csname bibfnamefont\endcsname\relax
  \def\bibfnamefont#1{#1}\fi
\expandafter\ifx\csname citenamefont\endcsname\relax
  \def\citenamefont#1{#1}\fi
\expandafter\ifx\csname url\endcsname\relax
  \def\url#1{\texttt{#1}}\fi
\expandafter\ifx\csname urlprefix\endcsname\relax\def\urlprefix{URL }\fi
\providecommand{\bibinfo}[2]{#2}
\providecommand{\eprint}[2][]{\url{#2}}

\bibitem[{\citenamefont{Cleland}(2002)}]{cleland:2002}
\bibinfo{author}{\bibfnamefont{A.}~\bibnamefont{Cleland}},
  \emph{\bibinfo{title}{Foundations of Nanomechanics}}, Advanced Texts in
  Physics (\bibinfo{publisher}{Springer}, \bibinfo{year}{2002}).

\bibitem[{\citenamefont{Blencowe}(2004)}]{blencowe:2004}
\bibinfo{author}{\bibfnamefont{M.}~\bibnamefont{Blencowe}},
  \bibinfo{journal}{Physics Reports} \textbf{\bibinfo{volume}{395}},
  \bibinfo{pages}{159 } (\bibinfo{year}{2004}).

\bibitem[{\citenamefont{Poot and van~der Zant}(2012)}]{poot:2012}
\bibinfo{author}{\bibfnamefont{M.}~\bibnamefont{Poot}} \bibnamefont{and}
  \bibinfo{author}{\bibfnamefont{H.~S.} \bibnamefont{van~der Zant}},
  \bibinfo{journal}{Physics Reports} \textbf{\bibinfo{volume}{511}},
  \bibinfo{pages}{273 } (\bibinfo{year}{2012}).

\bibitem[{\citenamefont{{Aspelmeyer} et~al.}(2013)\citenamefont{{Aspelmeyer},
  {Kippenberg}, and {Marquardt}}}]{aspelmeyer:2013}
\bibinfo{author}{\bibfnamefont{M.}~\bibnamefont{{Aspelmeyer}}},
  \bibinfo{author}{\bibfnamefont{T.~J.} \bibnamefont{{Kippenberg}}},
  \bibnamefont{and}
  \bibinfo{author}{\bibfnamefont{F.}~\bibnamefont{{Marquardt}}},
  \bibinfo{journal}{arXiv:1303.0733}  (\bibinfo{year}{2013}).

\bibitem[{\citenamefont{O'Connell et~al.}(2010)\citenamefont{O'Connell,
  Hofheinz, Ansmann, Bialczak, Lenander, Lucero, Neeley, Sank, Wang, Weides
  et~al.}}]{oconnell:2010}
\bibinfo{author}{\bibfnamefont{A.~D.} \bibnamefont{O'Connell}},
  \bibinfo{author}{\bibfnamefont{M.}~\bibnamefont{Hofheinz}},
  \bibinfo{author}{\bibfnamefont{M.}~\bibnamefont{Ansmann}},
  \bibinfo{author}{\bibfnamefont{R.~C.} \bibnamefont{Bialczak}},
  \bibinfo{author}{\bibfnamefont{M.}~\bibnamefont{Lenander}},
  \bibinfo{author}{\bibfnamefont{E.}~\bibnamefont{Lucero}},
  \bibinfo{author}{\bibfnamefont{M.}~\bibnamefont{Neeley}},
  \bibinfo{author}{\bibfnamefont{D.}~\bibnamefont{Sank}},
  \bibinfo{author}{\bibfnamefont{H.}~\bibnamefont{Wang}},
  \bibinfo{author}{\bibfnamefont{M.}~\bibnamefont{Weides}},
  \bibnamefont{et~al.}, \bibinfo{journal}{Nature}
  \textbf{\bibinfo{volume}{464}}, \bibinfo{pages}{697} (\bibinfo{year}{2010}).

\bibitem[{\citenamefont{Teufel et~al.}(2011)\citenamefont{Teufel, Donner, Li,
  Harlow, Allman, Cicak, Sirois, Whittaker, Lehnert, and
  Simmonds}}]{teufel:2011:2}
\bibinfo{author}{\bibfnamefont{J.~D.} \bibnamefont{Teufel}},
  \bibinfo{author}{\bibfnamefont{T.}~\bibnamefont{Donner}},
  \bibinfo{author}{\bibfnamefont{D.}~\bibnamefont{Li}},
  \bibinfo{author}{\bibfnamefont{J.~W.} \bibnamefont{Harlow}},
  \bibinfo{author}{\bibfnamefont{M.~S.} \bibnamefont{Allman}},
  \bibinfo{author}{\bibfnamefont{K.}~\bibnamefont{Cicak}},
  \bibinfo{author}{\bibfnamefont{A.~J.} \bibnamefont{Sirois}},
  \bibinfo{author}{\bibfnamefont{J.~D.} \bibnamefont{Whittaker}},
  \bibinfo{author}{\bibfnamefont{K.~W.} \bibnamefont{Lehnert}},
  \bibnamefont{and} \bibinfo{author}{\bibfnamefont{R.~W.}
  \bibnamefont{Simmonds}}, \bibinfo{journal}{Nature}
  \textbf{\bibinfo{volume}{475}}, \bibinfo{pages}{359} (\bibinfo{year}{2011}).

\bibitem[{\citenamefont{Chan et~al.}(2011)\citenamefont{Chan, Alegre,
  Safavi-Naeini, Hill, Krause, Groblacher, Aspelmeyer, and
  Painter}}]{chan:2011}
\bibinfo{author}{\bibfnamefont{J.}~\bibnamefont{Chan}},
  \bibinfo{author}{\bibfnamefont{T.~P.~M.} \bibnamefont{Alegre}},
  \bibinfo{author}{\bibfnamefont{A.~H.} \bibnamefont{Safavi-Naeini}},
  \bibinfo{author}{\bibfnamefont{J.~T.} \bibnamefont{Hill}},
  \bibinfo{author}{\bibfnamefont{A.}~\bibnamefont{Krause}},
  \bibinfo{author}{\bibfnamefont{S.}~\bibnamefont{Groblacher}},
  \bibinfo{author}{\bibfnamefont{M.}~\bibnamefont{Aspelmeyer}},
  \bibnamefont{and} \bibinfo{author}{\bibfnamefont{O.}~\bibnamefont{Painter}},
  \bibinfo{journal}{Nature} \textbf{\bibinfo{volume}{478}}, \bibinfo{pages}{89}
  (\bibinfo{year}{2011}).

\bibitem[{\citenamefont{Regal et~al.}(2008)\citenamefont{Regal, Teufel, and
  Lehnert}}]{regal:2008}
\bibinfo{author}{\bibfnamefont{C.~A.} \bibnamefont{Regal}},
  \bibinfo{author}{\bibfnamefont{J.~D.} \bibnamefont{Teufel}},
  \bibnamefont{and} \bibinfo{author}{\bibfnamefont{K.~W.}
  \bibnamefont{Lehnert}}, \bibinfo{journal}{Nat Phys}
  \textbf{\bibinfo{volume}{4}}, \bibinfo{pages}{555} (\bibinfo{year}{2008}).

\bibitem[{\citenamefont{Sun et~al.}(2006)\citenamefont{Sun, Wei, Liu, and
  Nori}}]{sun:2006}
\bibinfo{author}{\bibfnamefont{C.~P.} \bibnamefont{Sun}},
  \bibinfo{author}{\bibfnamefont{L.~F.} \bibnamefont{Wei}},
  \bibinfo{author}{\bibfnamefont{Y.-x.} \bibnamefont{Liu}}, \bibnamefont{and}
  \bibinfo{author}{\bibfnamefont{F.}~\bibnamefont{Nori}},
  \bibinfo{journal}{Phys. Rev. A} \textbf{\bibinfo{volume}{73}},
  \bibinfo{pages}{022318} (\bibinfo{year}{2006}).

\bibitem[{\citenamefont{Wallquist et~al.}(2009)\citenamefont{Wallquist,
  Hammerer, Rabl, Lukin, and Zoller}}]{wallquist:2009}
\bibinfo{author}{\bibfnamefont{M.}~\bibnamefont{Wallquist}},
  \bibinfo{author}{\bibfnamefont{K.}~\bibnamefont{Hammerer}},
  \bibinfo{author}{\bibfnamefont{P.}~\bibnamefont{Rabl}},
  \bibinfo{author}{\bibfnamefont{M.}~\bibnamefont{Lukin}}, \bibnamefont{and}
  \bibinfo{author}{\bibfnamefont{P.}~\bibnamefont{Zoller}},
  \bibinfo{journal}{Physica Scripta} \textbf{\bibinfo{volume}{2009}},
  \bibinfo{pages}{014001} (\bibinfo{year}{2009}).

\bibitem[{\citenamefont{Safavi-Naeini and Painter}(2011)}]{safavi-naeini:2011}
\bibinfo{author}{\bibfnamefont{A.~H.} \bibnamefont{Safavi-Naeini}}
  \bibnamefont{and} \bibinfo{author}{\bibfnamefont{O.}~\bibnamefont{Painter}},
  \bibinfo{journal}{New Journal of Physics} \textbf{\bibinfo{volume}{13}},
  \bibinfo{pages}{013017} (\bibinfo{year}{2011}).

\bibitem[{\citenamefont{Xiang et~al.}(2013)\citenamefont{Xiang, Ashhab, You,
  and Nori}}]{xiang:2013}
\bibinfo{author}{\bibfnamefont{Z.-L.} \bibnamefont{Xiang}},
  \bibinfo{author}{\bibfnamefont{S.}~\bibnamefont{Ashhab}},
  \bibinfo{author}{\bibfnamefont{J.~Q.} \bibnamefont{You}}, \bibnamefont{and}
  \bibinfo{author}{\bibfnamefont{F.}~\bibnamefont{Nori}},
  \bibinfo{journal}{Rev. Mod. Phys.} \textbf{\bibinfo{volume}{85}},
  \bibinfo{pages}{623} (\bibinfo{year}{2013}).

\bibitem[{\citenamefont{Bochmann et~al.}(2013)\citenamefont{Bochmann,
  Vainsencher, Awschalom, and Cleland}}]{bochmann:2013}
\bibinfo{author}{\bibfnamefont{J.}~\bibnamefont{Bochmann}},
  \bibinfo{author}{\bibfnamefont{A.}~\bibnamefont{Vainsencher}},
  \bibinfo{author}{\bibfnamefont{D.~D.} \bibnamefont{Awschalom}},
  \bibnamefont{and} \bibinfo{author}{\bibfnamefont{A.~N.}
  \bibnamefont{Cleland}}, \bibinfo{journal}{Nat Phys}
  \textbf{\bibinfo{volume}{9}}, \bibinfo{pages}{712} (\bibinfo{year}{2013}).

\bibitem[{\citenamefont{Rips and Hartmann}(2013)}]{rips:2013}
\bibinfo{author}{\bibfnamefont{S.}~\bibnamefont{Rips}} \bibnamefont{and}
  \bibinfo{author}{\bibfnamefont{M.~J.} \bibnamefont{Hartmann}},
  \bibinfo{journal}{Phys. Rev. Lett.} \textbf{\bibinfo{volume}{110}},
  \bibinfo{pages}{120503} (\bibinfo{year}{2013}).

\bibitem[{\citenamefont{Tian et~al.}(2008)\citenamefont{Tian, Allman, and
  Simmonds}}]{tian:2008}
\bibinfo{author}{\bibfnamefont{L.}~\bibnamefont{Tian}},
  \bibinfo{author}{\bibfnamefont{M.~S.} \bibnamefont{Allman}},
  \bibnamefont{and} \bibinfo{author}{\bibfnamefont{R.~W.}
  \bibnamefont{Simmonds}}, \bibinfo{journal}{New Journal of Physics}
  \textbf{\bibinfo{volume}{10}}, \bibinfo{pages}{115001}
  (\bibinfo{year}{2008}).

\bibitem[{\citenamefont{J\"ahne et~al.}(2009)\citenamefont{J\"ahne, Genes,
  Hammerer, Wallquist, Polzik, and Zoller}}]{jahne:2009}
\bibinfo{author}{\bibfnamefont{K.}~\bibnamefont{J\"ahne}},
  \bibinfo{author}{\bibfnamefont{C.}~\bibnamefont{Genes}},
  \bibinfo{author}{\bibfnamefont{K.}~\bibnamefont{Hammerer}},
  \bibinfo{author}{\bibfnamefont{M.}~\bibnamefont{Wallquist}},
  \bibinfo{author}{\bibfnamefont{E.~S.} \bibnamefont{Polzik}},
  \bibnamefont{and} \bibinfo{author}{\bibfnamefont{P.}~\bibnamefont{Zoller}},
  \bibinfo{journal}{Phys. Rev. A} \textbf{\bibinfo{volume}{79}},
  \bibinfo{pages}{063819} (\bibinfo{year}{2009}).

\bibitem[{\citenamefont{Stannigel et~al.}(2011)\citenamefont{Stannigel, Rabl,
  S\o{}rensen, Lukin, and Zoller}}]{stannigel:2011}
\bibinfo{author}{\bibfnamefont{K.}~\bibnamefont{Stannigel}},
  \bibinfo{author}{\bibfnamefont{P.}~\bibnamefont{Rabl}},
  \bibinfo{author}{\bibfnamefont{A.~S.} \bibnamefont{S\o{}rensen}},
  \bibinfo{author}{\bibfnamefont{M.~D.} \bibnamefont{Lukin}}, \bibnamefont{and}
  \bibinfo{author}{\bibfnamefont{P.}~\bibnamefont{Zoller}},
  \bibinfo{journal}{Phys. Rev. A} \textbf{\bibinfo{volume}{84}},
  \bibinfo{pages}{042341} (\bibinfo{year}{2011}).

\bibitem[{\citenamefont{Wang and Clerk}(2013)}]{wang:2013}
\bibinfo{author}{\bibfnamefont{Y.-D.} \bibnamefont{Wang}} \bibnamefont{and}
  \bibinfo{author}{\bibfnamefont{A.~A.} \bibnamefont{Clerk}},
  \bibinfo{journal}{Phys. Rev. Lett.} \textbf{\bibinfo{volume}{110}},
  \bibinfo{pages}{253601} (\bibinfo{year}{2013}).

\bibitem[{\citenamefont{Westra et~al.}(2010)\citenamefont{Westra, Poot, van~der
  Zant, and Venstra}}]{westra:2010}
\bibinfo{author}{\bibfnamefont{H.~J.~R.} \bibnamefont{Westra}},
  \bibinfo{author}{\bibfnamefont{M.}~\bibnamefont{Poot}},
  \bibinfo{author}{\bibfnamefont{H.~S.~J.} \bibnamefont{van~der Zant}},
  \bibnamefont{and} \bibinfo{author}{\bibfnamefont{W.~J.}
  \bibnamefont{Venstra}}, \bibinfo{journal}{Phys. Rev. Lett.}
  \textbf{\bibinfo{volume}{105}}, \bibinfo{pages}{117205}
  (\bibinfo{year}{2010}).

\bibitem[{\citenamefont{Lulla et~al.}(2012)\citenamefont{Lulla, Cousins,
  Venkatesan, Patton, Armour, Mellor, and Owers-Bradley}}]{lulla:2012}
\bibinfo{author}{\bibfnamefont{K.~J.} \bibnamefont{Lulla}},
  \bibinfo{author}{\bibfnamefont{R.~B.} \bibnamefont{Cousins}},
  \bibinfo{author}{\bibfnamefont{A.}~\bibnamefont{Venkatesan}},
  \bibinfo{author}{\bibfnamefont{M.~J.} \bibnamefont{Patton}},
  \bibinfo{author}{\bibfnamefont{A.~D.} \bibnamefont{Armour}},
  \bibinfo{author}{\bibfnamefont{C.~J.} \bibnamefont{Mellor}},
  \bibnamefont{and} \bibinfo{author}{\bibfnamefont{J.~R.}
  \bibnamefont{Owers-Bradley}}, \bibinfo{journal}{New Journal of Physics}
  \textbf{\bibinfo{volume}{14}}, \bibinfo{pages}{113040}
  (\bibinfo{year}{2012}).

\bibitem[{\citenamefont{Khan et~al.}(2013)\citenamefont{Khan, Massel, and
  Heikkil\"a}}]{khan:2013}
\bibinfo{author}{\bibfnamefont{R.}~\bibnamefont{Khan}},
  \bibinfo{author}{\bibfnamefont{F.}~\bibnamefont{Massel}}, \bibnamefont{and}
  \bibinfo{author}{\bibfnamefont{T.~T.} \bibnamefont{Heikkil\"a}},
  \bibinfo{journal}{Phys. Rev. B} \textbf{\bibinfo{volume}{87}},
  \bibinfo{pages}{235406} (\bibinfo{year}{2013}).

\bibitem[{\citenamefont{Yamaguchi and Mahboob}(2013)}]{yamaguchi:2013}
\bibinfo{author}{\bibfnamefont{H.}~\bibnamefont{Yamaguchi}} \bibnamefont{and}
  \bibinfo{author}{\bibfnamefont{I.}~\bibnamefont{Mahboob}},
  \bibinfo{journal}{New Journal of Physics} \textbf{\bibinfo{volume}{15}},
  \bibinfo{pages}{015023} (\bibinfo{year}{2013}).

\bibitem[{\citenamefont{Clerk et~al.}(2008)\citenamefont{Clerk, Marquardt, and
  Jacobs}}]{clerk:2008}
\bibinfo{author}{\bibfnamefont{A.~A.} \bibnamefont{Clerk}},
  \bibinfo{author}{\bibfnamefont{F.}~\bibnamefont{Marquardt}},
  \bibnamefont{and} \bibinfo{author}{\bibfnamefont{K.}~\bibnamefont{Jacobs}},
  \bibinfo{journal}{New Journal of Physics} \textbf{\bibinfo{volume}{10}},
  \bibinfo{pages}{095010} (\bibinfo{year}{2008}).

\bibitem[{\citenamefont{Liao and Law}(2011)}]{liao:2011}
\bibinfo{author}{\bibfnamefont{J.-Q.} \bibnamefont{Liao}} \bibnamefont{and}
  \bibinfo{author}{\bibfnamefont{C.~K.} \bibnamefont{Law}},
  \bibinfo{journal}{Phys. Rev. A} \textbf{\bibinfo{volume}{83}},
  \bibinfo{pages}{033820} (\bibinfo{year}{2011}).

\bibitem[{\citenamefont{Palomaki
  et~al.}(2013{\natexlab{a}})\citenamefont{Palomaki, Harlow, Teufel, Simmonds,
  and Lehnert}}]{palomaki:2013}
\bibinfo{author}{\bibfnamefont{T.~A.} \bibnamefont{Palomaki}},
  \bibinfo{author}{\bibfnamefont{J.~W.} \bibnamefont{Harlow}},
  \bibinfo{author}{\bibfnamefont{J.~D.} \bibnamefont{Teufel}},
  \bibinfo{author}{\bibfnamefont{R.~W.} \bibnamefont{Simmonds}},
  \bibnamefont{and} \bibinfo{author}{\bibfnamefont{K.~W.}
  \bibnamefont{Lehnert}}, \bibinfo{journal}{Nature}
  \textbf{\bibinfo{volume}{495}}, \bibinfo{pages}{210}
  (\bibinfo{year}{2013}{\natexlab{a}}).

\bibitem[{\citenamefont{Palomaki
  et~al.}(2013{\natexlab{b}})\citenamefont{Palomaki, Teufel, Simmonds, and
  Lehnert}}]{palomaki:2013:2}
\bibinfo{author}{\bibfnamefont{T.~A.} \bibnamefont{Palomaki}},
  \bibinfo{author}{\bibfnamefont{J.~D.} \bibnamefont{Teufel}},
  \bibinfo{author}{\bibfnamefont{R.~W.} \bibnamefont{Simmonds}},
  \bibnamefont{and} \bibinfo{author}{\bibfnamefont{K.~W.}
  \bibnamefont{Lehnert}}, \bibinfo{journal}{Science}
  \textbf{\bibinfo{volume}{8}}, \bibinfo{pages}{710}
  (\bibinfo{year}{2013}{\natexlab{b}}).

\bibitem[{\citenamefont{Lemonde et~al.}(2013)\citenamefont{Lemonde, Didier, and
  Clerk}}]{lemode:2013}
\bibinfo{author}{\bibfnamefont{M.-A.} \bibnamefont{Lemonde}},
  \bibinfo{author}{\bibfnamefont{N.}~\bibnamefont{Didier}}, \bibnamefont{and}
  \bibinfo{author}{\bibfnamefont{A.~A.} \bibnamefont{Clerk}},
  \bibinfo{journal}{Phys. Rev. Lett.} \textbf{\bibinfo{volume}{111}},
  \bibinfo{pages}{053602} (\bibinfo{year}{2013}).

\bibitem[{\citenamefont{Qian et~al.}(2012)\citenamefont{Qian, Clerk, Hammerer,
  and Marquardt}}]{qian:2012}
\bibinfo{author}{\bibfnamefont{J.}~\bibnamefont{Qian}},
  \bibinfo{author}{\bibfnamefont{A.~A.} \bibnamefont{Clerk}},
  \bibinfo{author}{\bibfnamefont{K.}~\bibnamefont{Hammerer}}, \bibnamefont{and}
  \bibinfo{author}{\bibfnamefont{F.}~\bibnamefont{Marquardt}},
  \bibinfo{journal}{Phys. Rev. Lett.} \textbf{\bibinfo{volume}{109}},
  \bibinfo{pages}{253601} (\bibinfo{year}{2012}).

\bibitem[{\citenamefont{Nation}(2013)}]{nation:2013}
\bibinfo{author}{\bibfnamefont{P.~D.} \bibnamefont{Nation}},
  \bibinfo{journal}{Phys. Rev. A} \textbf{\bibinfo{volume}{88}},
  \bibinfo{pages}{053828} (\bibinfo{year}{2013}).

\bibitem[{\citenamefont{{L{\"o}rch} et~al.}(2013)\citenamefont{{L{\"o}rch},
  {Qian}, {Clerk}, {Marquardt}, and {Hammerer}}}]{lorch:2012}
\bibinfo{author}{\bibfnamefont{N.}~\bibnamefont{{L{\"o}rch}}},
  \bibinfo{author}{\bibfnamefont{J.}~\bibnamefont{{Qian}}},
  \bibinfo{author}{\bibfnamefont{A.}~\bibnamefont{{Clerk}}},
  \bibinfo{author}{\bibfnamefont{F.}~\bibnamefont{{Marquardt}}},
  \bibnamefont{and}
  \bibinfo{author}{\bibfnamefont{K.}~\bibnamefont{{Hammerer}}},
  \bibinfo{journal}{arXiv:1310.1298}  (\bibinfo{year}{2013}).

\bibitem[{\citenamefont{Rips et~al.}(2012)\citenamefont{Rips, Kiffner,
  Wilson-Rae, and Hartmann}}]{rips:2012}
\bibinfo{author}{\bibfnamefont{S.}~\bibnamefont{Rips}},
  \bibinfo{author}{\bibfnamefont{M.}~\bibnamefont{Kiffner}},
  \bibinfo{author}{\bibfnamefont{I.}~\bibnamefont{Wilson-Rae}},
  \bibnamefont{and} \bibinfo{author}{\bibfnamefont{M.~J.}
  \bibnamefont{Hartmann}}, \bibinfo{journal}{New Journal of Physics}
  \textbf{\bibinfo{volume}{14}}, \bibinfo{pages}{023042}
  (\bibinfo{year}{2012}).

\bibitem[{\citenamefont{Tian}(2005)}]{tian:2005}
\bibinfo{author}{\bibfnamefont{L.}~\bibnamefont{Tian}}, \bibinfo{journal}{Phys.
  Rev. B} \textbf{\bibinfo{volume}{72}}, \bibinfo{pages}{195411}
  (\bibinfo{year}{2005}).

\bibitem[{\citenamefont{Voje et~al.}(2012)\citenamefont{Voje, Kinaret, and
  Isacsson}}]{voje:2012}
\bibinfo{author}{\bibfnamefont{A.}~\bibnamefont{Voje}},
  \bibinfo{author}{\bibfnamefont{J.~M.} \bibnamefont{Kinaret}},
  \bibnamefont{and} \bibinfo{author}{\bibfnamefont{A.}~\bibnamefont{Isacsson}},
  \bibinfo{journal}{Phys. Rev. B} \textbf{\bibinfo{volume}{85}},
  \bibinfo{pages}{205415} (\bibinfo{year}{2012}).

\bibitem[{\citenamefont{Xue et~al.}(2007)\citenamefont{Xue, Liu, Sun, and
  Nori}}]{fei:2007}
\bibinfo{author}{\bibfnamefont{F.}~\bibnamefont{Xue}},
  \bibinfo{author}{\bibfnamefont{Y.-x.} \bibnamefont{Liu}},
  \bibinfo{author}{\bibfnamefont{C.~P.} \bibnamefont{Sun}}, \bibnamefont{and}
  \bibinfo{author}{\bibfnamefont{F.}~\bibnamefont{Nori}},
  \bibinfo{journal}{Phys. Rev. B} \textbf{\bibinfo{volume}{76}},
  \bibinfo{pages}{064305} (\bibinfo{year}{2007}).

\bibitem[{\citenamefont{Cohen and Di~Ventra}(2013)}]{cohen:2013}
\bibinfo{author}{\bibfnamefont{G.~Z.} \bibnamefont{Cohen}} \bibnamefont{and}
  \bibinfo{author}{\bibfnamefont{M.}~\bibnamefont{Di~Ventra}},
  \bibinfo{journal}{Phys. Rev. B} \textbf{\bibinfo{volume}{87}},
  \bibinfo{pages}{014513} (\bibinfo{year}{2013}).

\bibitem[{\citenamefont{Tan et~al.}(2013)\citenamefont{Tan, Li, and
  Meystre}}]{huatang:2013}
\bibinfo{author}{\bibfnamefont{H.}~\bibnamefont{Tan}},
  \bibinfo{author}{\bibfnamefont{G.}~\bibnamefont{Li}}, \bibnamefont{and}
  \bibinfo{author}{\bibfnamefont{P.}~\bibnamefont{Meystre}},
  \bibinfo{journal}{Phys. Rev. A} \textbf{\bibinfo{volume}{87}},
  \bibinfo{pages}{033829} (\bibinfo{year}{2013}).

\bibitem[{\citenamefont{Xu et~al.}(2013)\citenamefont{Xu, Zhao, and
  Liu}}]{xu:2013}
\bibinfo{author}{\bibfnamefont{X.-W.} \bibnamefont{Xu}},
  \bibinfo{author}{\bibfnamefont{Y.-J.} \bibnamefont{Zhao}}, \bibnamefont{and}
  \bibinfo{author}{\bibfnamefont{Y.-x.} \bibnamefont{Liu}},
  \bibinfo{journal}{Phys. Rev. A} \textbf{\bibinfo{volume}{88}},
  \bibinfo{pages}{022325} (\bibinfo{year}{2013}).

\bibitem[{\citenamefont{Suh et~al.}(2010)\citenamefont{Suh, LaHaye, Echternach,
  Schwab, and Roukes}}]{junho:2010}
\bibinfo{author}{\bibfnamefont{J.}~\bibnamefont{Suh}},
  \bibinfo{author}{\bibfnamefont{M.~D.} \bibnamefont{LaHaye}},
  \bibinfo{author}{\bibfnamefont{P.~M.} \bibnamefont{Echternach}},
  \bibinfo{author}{\bibfnamefont{K.~C.} \bibnamefont{Schwab}},
  \bibnamefont{and} \bibinfo{author}{\bibfnamefont{M.~L.}
  \bibnamefont{Roukes}}, \bibinfo{journal}{Nano Letters}
  \textbf{\bibinfo{volume}{10}}, \bibinfo{pages}{3990} (\bibinfo{year}{2010}).

\bibitem[{\citenamefont{Massel et~al.}(2011)\citenamefont{Massel, Heikkila,
  Pirkkalainen, Cho, Saloniemi, Hakonen, and Sillanpaa}}]{massel:2011}
\bibinfo{author}{\bibfnamefont{F.}~\bibnamefont{Massel}},
  \bibinfo{author}{\bibfnamefont{T.~T.} \bibnamefont{Heikkila}},
  \bibinfo{author}{\bibfnamefont{J.-M.} \bibnamefont{Pirkkalainen}},
  \bibinfo{author}{\bibfnamefont{S.~U.} \bibnamefont{Cho}},
  \bibinfo{author}{\bibfnamefont{H.}~\bibnamefont{Saloniemi}},
  \bibinfo{author}{\bibfnamefont{P.~J.} \bibnamefont{Hakonen}},
  \bibnamefont{and} \bibinfo{author}{\bibfnamefont{M.~A.}
  \bibnamefont{Sillanpaa}}, \bibinfo{journal}{Nature}
  \textbf{\bibinfo{volume}{480}}, \bibinfo{pages}{351} (\bibinfo{year}{2011}).

\bibitem[{\citenamefont{Eichler
  et~al.}(2011{\natexlab{a}})\citenamefont{Eichler, Moser, Chaste, Zdrojek,
  Wilson-Rae, and Bachtold}}]{eichler:2011}
\bibinfo{author}{\bibfnamefont{A.}~\bibnamefont{Eichler}},
  \bibinfo{author}{\bibfnamefont{J.}~\bibnamefont{Moser}},
  \bibinfo{author}{\bibfnamefont{J.}~\bibnamefont{Chaste}},
  \bibinfo{author}{\bibfnamefont{M.}~\bibnamefont{Zdrojek}},
  \bibinfo{author}{\bibfnamefont{I.}~\bibnamefont{Wilson-Rae}},
  \bibnamefont{and} \bibinfo{author}{\bibfnamefont{A.}~\bibnamefont{Bachtold}},
  \bibinfo{journal}{Nat. Nano.} \textbf{\bibinfo{volume}{6}},
  \bibinfo{pages}{339} (\bibinfo{year}{2011}{\natexlab{a}}).

\bibitem[{\citenamefont{Voje et~al.}(2013)\citenamefont{Voje, Isacsson, and
  Croy}}]{voje:2013}
\bibinfo{author}{\bibfnamefont{A.}~\bibnamefont{Voje}},
  \bibinfo{author}{\bibfnamefont{A.}~\bibnamefont{Isacsson}}, \bibnamefont{and}
  \bibinfo{author}{\bibfnamefont{A.}~\bibnamefont{Croy}},
  \bibinfo{journal}{Phys. Rev. A} \textbf{\bibinfo{volume}{88}},
  \bibinfo{pages}{022309} (\bibinfo{year}{2013}).

\bibitem[{\citenamefont{{Jacobs}}(2012)}]{jacobs:2012}
\bibinfo{author}{\bibfnamefont{K.}~\bibnamefont{{Jacobs}}},
  \bibinfo{journal}{arXiv:1209.2499}  (\bibinfo{year}{2012}).

\bibitem[{\citenamefont{Mahboob et~al.}(2013)\citenamefont{Mahboob, Nishiguchi,
  Fujiwara, and Yamaguchi}}]{mahboob:2013}
\bibinfo{author}{\bibfnamefont{I.}~\bibnamefont{Mahboob}},
  \bibinfo{author}{\bibfnamefont{K.}~\bibnamefont{Nishiguchi}},
  \bibinfo{author}{\bibfnamefont{A.}~\bibnamefont{Fujiwara}}, \bibnamefont{and}
  \bibinfo{author}{\bibfnamefont{H.}~\bibnamefont{Yamaguchi}},
  \bibinfo{journal}{Phys. Rev. Lett.} \textbf{\bibinfo{volume}{110}},
  \bibinfo{pages}{127202} (\bibinfo{year}{2013}).

\bibitem[{\citenamefont{Kheruntsyan and Petrosyan}(2000)}]{kheruntsyan:2000}
\bibinfo{author}{\bibfnamefont{K.~V.} \bibnamefont{Kheruntsyan}}
  \bibnamefont{and} \bibinfo{author}{\bibfnamefont{K.~G.}
  \bibnamefont{Petrosyan}}, \bibinfo{journal}{Phys. Rev. A}
  \textbf{\bibinfo{volume}{62}}, \bibinfo{pages}{015801}
  (\bibinfo{year}{2000}).

\bibitem[{\citenamefont{McNeil and Gardiner}(1983)}]{mcneil:1983}
\bibinfo{author}{\bibfnamefont{K.~J.} \bibnamefont{McNeil}} \bibnamefont{and}
  \bibinfo{author}{\bibfnamefont{C.~W.} \bibnamefont{Gardiner}},
  \bibinfo{journal}{Phys. Rev.} \textbf{\bibinfo{volume}{28}},
  \bibinfo{pages}{1560} (\bibinfo{year}{1983}).

\bibitem[{\citenamefont{Munro}(1999)}]{munro:1999}
\bibinfo{author}{\bibfnamefont{W.~J.} \bibnamefont{Munro}},
  \bibinfo{journal}{Phys. Rev. A} \textbf{\bibinfo{volume}{59}},
  \bibinfo{pages}{4197} (\bibinfo{year}{1999}).

\bibitem[{\citenamefont{Reid and Krippner}(1993)}]{reid:1993}
\bibinfo{author}{\bibfnamefont{M.~D.} \bibnamefont{Reid}} \bibnamefont{and}
  \bibinfo{author}{\bibfnamefont{L.}~\bibnamefont{Krippner}},
  \bibinfo{journal}{Phys. Rev. A} \textbf{\bibinfo{volume}{47}},
  \bibinfo{pages}{552} (\bibinfo{year}{1993}).

\bibitem[{\citenamefont{Wiseman and Milburn}(1993)}]{wiseman:1993}
\bibinfo{author}{\bibfnamefont{H.~M.} \bibnamefont{Wiseman}} \bibnamefont{and}
  \bibinfo{author}{\bibfnamefont{G.~J.} \bibnamefont{Milburn}},
  \bibinfo{journal}{Phys. Rev. A} \textbf{\bibinfo{volume}{47}},
  \bibinfo{pages}{642} (\bibinfo{year}{1993}).

\bibitem[{\citenamefont{Reid and Drummond}(1988)}]{reid:1988}
\bibinfo{author}{\bibfnamefont{M.~D.} \bibnamefont{Reid}} \bibnamefont{and}
  \bibinfo{author}{\bibfnamefont{P.~D.} \bibnamefont{Drummond}},
  \bibinfo{journal}{Phys. Rev. Lett.} \textbf{\bibinfo{volume}{60}},
  \bibinfo{pages}{2731} (\bibinfo{year}{1988}).

\bibitem[{\citenamefont{Marian et~al.}(2003)\citenamefont{Marian, Marian, and
  Scutaru}}]{marian:2003}
\bibinfo{author}{\bibfnamefont{P.}~\bibnamefont{Marian}},
  \bibinfo{author}{\bibfnamefont{T.~A.} \bibnamefont{Marian}},
  \bibnamefont{and} \bibinfo{author}{\bibfnamefont{H.}~\bibnamefont{Scutaru}},
  \bibinfo{journal}{Phys. Rev. A} \textbf{\bibinfo{volume}{68}},
  \bibinfo{pages}{062309} (\bibinfo{year}{2003}).

\bibitem[{\citenamefont{Gilchrist et~al.}(1998)\citenamefont{Gilchrist, Deuar,
  and Reid}}]{gilchrist:1998}
\bibinfo{author}{\bibfnamefont{A.}~\bibnamefont{Gilchrist}},
  \bibinfo{author}{\bibfnamefont{P.}~\bibnamefont{Deuar}}, \bibnamefont{and}
  \bibinfo{author}{\bibfnamefont{M.~D.} \bibnamefont{Reid}},
  \bibinfo{journal}{Phys. Rev. Lett.} \textbf{\bibinfo{volume}{80}},
  \bibinfo{pages}{3169} (\bibinfo{year}{1998}).

\bibitem[{\citenamefont{Adesso and Illuminati}(2007)}]{adesso:2007}
\bibinfo{author}{\bibfnamefont{G.}~\bibnamefont{Adesso}} \bibnamefont{and}
  \bibinfo{author}{\bibfnamefont{F.}~\bibnamefont{Illuminati}},
  \bibinfo{journal}{Journal of Physics A: Mathematical and Theoretical}
  \textbf{\bibinfo{volume}{40}}, \bibinfo{pages}{7821} (\bibinfo{year}{2007}).

\bibitem[{\citenamefont{Fink et~al.}(2008)\citenamefont{Fink, G\"{o}ppl, Baur,
  Bianchetti, Leek, Blais, and Wallraff}}]{fink:2008}
\bibinfo{author}{\bibfnamefont{J.~M.} \bibnamefont{Fink}},
  \bibinfo{author}{\bibfnamefont{M.}~\bibnamefont{G\"{o}ppl}},
  \bibinfo{author}{\bibfnamefont{M.}~\bibnamefont{Baur}},
  \bibinfo{author}{\bibfnamefont{R.}~\bibnamefont{Bianchetti}},
  \bibinfo{author}{\bibfnamefont{P.~J.} \bibnamefont{Leek}},
  \bibinfo{author}{\bibfnamefont{A.}~\bibnamefont{Blais}}, \bibnamefont{and}
  \bibinfo{author}{\bibfnamefont{A.}~\bibnamefont{Wallraff}},
  \bibinfo{journal}{Nature} \textbf{\bibinfo{volume}{454}},
  \bibinfo{pages}{315} (\bibinfo{year}{2008}).

\bibitem[{\citenamefont{Schuster et~al.}(2008)\citenamefont{Schuster, Kubanek,
  Fuhrmanek, Puppe, Pinkse, Murr, and Rempe}}]{schuster:2008}
\bibinfo{author}{\bibfnamefont{I.}~\bibnamefont{Schuster}},
  \bibinfo{author}{\bibfnamefont{A.}~\bibnamefont{Kubanek}},
  \bibinfo{author}{\bibfnamefont{A.}~\bibnamefont{Fuhrmanek}},
  \bibinfo{author}{\bibfnamefont{T.}~\bibnamefont{Puppe}},
  \bibinfo{author}{\bibfnamefont{P.~W.~H.} \bibnamefont{Pinkse}},
  \bibinfo{author}{\bibfnamefont{K.}~\bibnamefont{Murr}}, \bibnamefont{and}
  \bibinfo{author}{\bibfnamefont{G.}~\bibnamefont{Rempe}},
  \bibinfo{journal}{Nature Physics} \textbf{\bibinfo{volume}{4}},
  \bibinfo{pages}{382} (\bibinfo{year}{2008}).

\bibitem[{\citenamefont{Miranowicz et~al.}(2010)\citenamefont{Miranowicz,
  Bartkowiak, Wang, Liu, and Nori}}]{miranowicz:2010}
\bibinfo{author}{\bibfnamefont{A.}~\bibnamefont{Miranowicz}},
  \bibinfo{author}{\bibfnamefont{M.}~\bibnamefont{Bartkowiak}},
  \bibinfo{author}{\bibfnamefont{X.}~\bibnamefont{Wang}},
  \bibinfo{author}{\bibfnamefont{Y.-x.} \bibnamefont{Liu}}, \bibnamefont{and}
  \bibinfo{author}{\bibfnamefont{F.}~\bibnamefont{Nori}},
  \bibinfo{journal}{Phys. Rev. A} \textbf{\bibinfo{volume}{82}},
  \bibinfo{pages}{013824} (\bibinfo{year}{2010}).

\bibitem[{\citenamefont{Simon}(2000)}]{simon:2000}
\bibinfo{author}{\bibfnamefont{R.}~\bibnamefont{Simon}},
  \bibinfo{journal}{Phys. Rev. Lett.} \textbf{\bibinfo{volume}{84}},
  \bibinfo{pages}{2726} (\bibinfo{year}{2000}).

\bibitem[{\citenamefont{Duan et~al.}(2000)\citenamefont{Duan, Giedke, Cirac,
  and Zoller}}]{duan:2000}
\bibinfo{author}{\bibfnamefont{L.-M.} \bibnamefont{Duan}},
  \bibinfo{author}{\bibfnamefont{G.}~\bibnamefont{Giedke}},
  \bibinfo{author}{\bibfnamefont{J.~I.} \bibnamefont{Cirac}}, \bibnamefont{and}
  \bibinfo{author}{\bibfnamefont{P.}~\bibnamefont{Zoller}},
  \bibinfo{journal}{Phys. Rev. Lett.} \textbf{\bibinfo{volume}{84}},
  \bibinfo{pages}{2722} (\bibinfo{year}{2000}).

\bibitem[{\citenamefont{{Brunner} et~al.}(2013)\citenamefont{{Brunner},
  {Cavalcanti}, {Pironio}, {Scarani}, and {Wehner}}}]{brunner:2013}
\bibinfo{author}{\bibfnamefont{N.}~\bibnamefont{{Brunner}}},
  \bibinfo{author}{\bibfnamefont{D.}~\bibnamefont{{Cavalcanti}}},
  \bibinfo{author}{\bibfnamefont{S.}~\bibnamefont{{Pironio}}},
  \bibinfo{author}{\bibfnamefont{V.}~\bibnamefont{{Scarani}}},
  \bibnamefont{and} \bibinfo{author}{\bibfnamefont{S.}~\bibnamefont{{Wehner}}},
  \bibinfo{journal}{arXiv:1303.2849}  (\bibinfo{year}{2013}).

\bibitem[{\citenamefont{Cavalcanti et~al.}(2007)\citenamefont{Cavalcanti,
  Foster, Reid, and Drummond}}]{cavalcanti:2007}
\bibinfo{author}{\bibfnamefont{E.~G.} \bibnamefont{Cavalcanti}},
  \bibinfo{author}{\bibfnamefont{C.~J.} \bibnamefont{Foster}},
  \bibinfo{author}{\bibfnamefont{M.~D.} \bibnamefont{Reid}}, \bibnamefont{and}
  \bibinfo{author}{\bibfnamefont{P.~D.} \bibnamefont{Drummond}},
  \bibinfo{journal}{Phys. Rev. Lett.} \textbf{\bibinfo{volume}{99}},
  \bibinfo{pages}{210405} (\bibinfo{year}{2007}).

\bibitem[{\citenamefont{Wenger et~al.}(2003)\citenamefont{Wenger, Hafezi,
  Grosshans, Tualle-Brouri, and Grangier}}]{wenger:2003}
\bibinfo{author}{\bibfnamefont{J.}~\bibnamefont{Wenger}},
  \bibinfo{author}{\bibfnamefont{M.}~\bibnamefont{Hafezi}},
  \bibinfo{author}{\bibfnamefont{F.}~\bibnamefont{Grosshans}},
  \bibinfo{author}{\bibfnamefont{R.}~\bibnamefont{Tualle-Brouri}},
  \bibnamefont{and} \bibinfo{author}{\bibfnamefont{P.}~\bibnamefont{Grangier}},
  \bibinfo{journal}{Phys. Rev. A} \textbf{\bibinfo{volume}{67}},
  \bibinfo{pages}{012105} (\bibinfo{year}{2003}).

\bibitem[{\citenamefont{Clauser and Horne}(1974)}]{clauser:1974}
\bibinfo{author}{\bibfnamefont{J.~F.} \bibnamefont{Clauser}} \bibnamefont{and}
  \bibinfo{author}{\bibfnamefont{M.~A.} \bibnamefont{Horne}},
  \bibinfo{journal}{Phys. Rev. D} \textbf{\bibinfo{volume}{10}},
  \bibinfo{pages}{526} (\bibinfo{year}{1974}).

\bibitem[{\citenamefont{Clauser et~al.}(1969)\citenamefont{Clauser, Horne,
  Shimony, and Holt}}]{clauser:1969}
\bibinfo{author}{\bibfnamefont{J.~F.} \bibnamefont{Clauser}},
  \bibinfo{author}{\bibfnamefont{M.~A.} \bibnamefont{Horne}},
  \bibinfo{author}{\bibfnamefont{A.}~\bibnamefont{Shimony}}, \bibnamefont{and}
  \bibinfo{author}{\bibfnamefont{R.~A.} \bibnamefont{Holt}},
  \bibinfo{journal}{Phys. Rev. Lett.} \textbf{\bibinfo{volume}{23}},
  \bibinfo{pages}{880} (\bibinfo{year}{1969}).

\bibitem[{\citenamefont{Grajcar et~al.}(2008)\citenamefont{Grajcar, Ashhab,
  Johansson, and Nori}}]{grajcar:2008}
\bibinfo{author}{\bibfnamefont{M.}~\bibnamefont{Grajcar}},
  \bibinfo{author}{\bibfnamefont{S.}~\bibnamefont{Ashhab}},
  \bibinfo{author}{\bibfnamefont{J.~R.} \bibnamefont{Johansson}},
  \bibnamefont{and} \bibinfo{author}{\bibfnamefont{F.}~\bibnamefont{Nori}},
  \bibinfo{journal}{Phys. Rev. B} \textbf{\bibinfo{volume}{78}},
  \bibinfo{pages}{035406} (\bibinfo{year}{2008}).

\bibitem[{\citenamefont{Sapmaz et~al.}(2003)\citenamefont{Sapmaz, Blanter,
  Gurevich, and van~der Zant}}]{sampaz:2003}
\bibinfo{author}{\bibfnamefont{S.}~\bibnamefont{Sapmaz}},
  \bibinfo{author}{\bibfnamefont{Y.~M.} \bibnamefont{Blanter}},
  \bibinfo{author}{\bibfnamefont{L.}~\bibnamefont{Gurevich}}, \bibnamefont{and}
  \bibinfo{author}{\bibfnamefont{H.~S.~J.} \bibnamefont{van~der Zant}},
  \bibinfo{journal}{Phys. Rev. B} \textbf{\bibinfo{volume}{67}},
  \bibinfo{pages}{235414} (\bibinfo{year}{2003}).

\bibitem[{\citenamefont{Eriksson et~al.}(2013)\citenamefont{Eriksson, Midtvedt,
  Croy, and Isacsson}}]{eriksson:2013}
\bibinfo{author}{\bibfnamefont{A.~M.} \bibnamefont{Eriksson}},
  \bibinfo{author}{\bibfnamefont{D.}~\bibnamefont{Midtvedt}},
  \bibinfo{author}{\bibfnamefont{A.}~\bibnamefont{Croy}}, \bibnamefont{and}
  \bibinfo{author}{\bibfnamefont{A.}~\bibnamefont{Isacsson}},
  \bibinfo{journal}{Nanotechnology} \textbf{\bibinfo{volume}{24}},
  \bibinfo{pages}{395702} (\bibinfo{year}{2013}).

\bibitem[{\citenamefont{Mahboob et~al.}(2012)\citenamefont{Mahboob, Nishiguchi,
  Okamoto, and Yamaguchi}}]{mahboob:2012}
\bibinfo{author}{\bibfnamefont{I.}~\bibnamefont{Mahboob}},
  \bibinfo{author}{\bibfnamefont{K.}~\bibnamefont{Nishiguchi}},
  \bibinfo{author}{\bibfnamefont{H.}~\bibnamefont{Okamoto}}, \bibnamefont{and}
  \bibinfo{author}{\bibfnamefont{H.}~\bibnamefont{Yamaguchi}},
  \bibinfo{journal}{Nat Phys} \textbf{\bibinfo{volume}{8}},
  \bibinfo{pages}{387} (\bibinfo{year}{2012}).

\bibitem[{\citenamefont{Paternostro et~al.}(2007)\citenamefont{Paternostro,
  Vitali, Gigan, Kim, Brukner, Eisert, and Aspelmeyer}}]{paternostro:2007}
\bibinfo{author}{\bibfnamefont{M.}~\bibnamefont{Paternostro}},
  \bibinfo{author}{\bibfnamefont{D.}~\bibnamefont{Vitali}},
  \bibinfo{author}{\bibfnamefont{S.}~\bibnamefont{Gigan}},
  \bibinfo{author}{\bibfnamefont{M.~S.} \bibnamefont{Kim}},
  \bibinfo{author}{\bibfnamefont{C.}~\bibnamefont{Brukner}},
  \bibinfo{author}{\bibfnamefont{J.}~\bibnamefont{Eisert}}, \bibnamefont{and}
  \bibinfo{author}{\bibfnamefont{M.}~\bibnamefont{Aspelmeyer}},
  \bibinfo{journal}{Phys. Rev. Lett.} \textbf{\bibinfo{volume}{99}},
  \bibinfo{pages}{250401} (\bibinfo{year}{2007}).

\bibitem[{\citenamefont{Eichler
  et~al.}(2011{\natexlab{b}})\citenamefont{Eichler, Bozyigit, Lang, Baur,
  Steffen, Fink, Filipp, and Wallraff}}]{eichler:2011:b}
\bibinfo{author}{\bibfnamefont{C.}~\bibnamefont{Eichler}},
  \bibinfo{author}{\bibfnamefont{D.}~\bibnamefont{Bozyigit}},
  \bibinfo{author}{\bibfnamefont{C.}~\bibnamefont{Lang}},
  \bibinfo{author}{\bibfnamefont{M.}~\bibnamefont{Baur}},
  \bibinfo{author}{\bibfnamefont{L.}~\bibnamefont{Steffen}},
  \bibinfo{author}{\bibfnamefont{J.~M.} \bibnamefont{Fink}},
  \bibinfo{author}{\bibfnamefont{S.}~\bibnamefont{Filipp}}, \bibnamefont{and}
  \bibinfo{author}{\bibfnamefont{A.}~\bibnamefont{Wallraff}},
  \bibinfo{journal}{Phys. Rev. Lett.} \textbf{\bibinfo{volume}{107}},
  \bibinfo{pages}{113601} (\bibinfo{year}{2011}{\natexlab{b}}).

\bibitem[{\citenamefont{Bergeal et~al.}(2012)\citenamefont{Bergeal, Schackert,
  Frunzio, and Devoret}}]{bergeal:2012}
\bibinfo{author}{\bibfnamefont{N.}~\bibnamefont{Bergeal}},
  \bibinfo{author}{\bibfnamefont{F.}~\bibnamefont{Schackert}},
  \bibinfo{author}{\bibfnamefont{L.}~\bibnamefont{Frunzio}}, \bibnamefont{and}
  \bibinfo{author}{\bibfnamefont{M.~H.} \bibnamefont{Devoret}},
  \bibinfo{journal}{Phys. Rev. Lett.} \textbf{\bibinfo{volume}{108}},
  \bibinfo{pages}{123902} (\bibinfo{year}{2012}).

\bibitem[{\citenamefont{Mallet et~al.}(2011)\citenamefont{Mallet,
  Castellanos-Beltran, Ku, Glancy, Knill, Irwin, Hilton, Vale, and
  Lehnert}}]{mallet:2011}
\bibinfo{author}{\bibfnamefont{F.}~\bibnamefont{Mallet}},
  \bibinfo{author}{\bibfnamefont{M.~A.} \bibnamefont{Castellanos-Beltran}},
  \bibinfo{author}{\bibfnamefont{H.~S.} \bibnamefont{Ku}},
  \bibinfo{author}{\bibfnamefont{S.}~\bibnamefont{Glancy}},
  \bibinfo{author}{\bibfnamefont{E.}~\bibnamefont{Knill}},
  \bibinfo{author}{\bibfnamefont{K.~D.} \bibnamefont{Irwin}},
  \bibinfo{author}{\bibfnamefont{G.~C.} \bibnamefont{Hilton}},
  \bibinfo{author}{\bibfnamefont{L.~R.} \bibnamefont{Vale}}, \bibnamefont{and}
  \bibinfo{author}{\bibfnamefont{K.~W.} \bibnamefont{Lehnert}},
  \bibinfo{journal}{Phys. Rev. Lett.} \textbf{\bibinfo{volume}{106}},
  \bibinfo{pages}{220502} (\bibinfo{year}{2011}).

\bibitem[{\citenamefont{Flurin et~al.}(2012)\citenamefont{Flurin, Roch, Mallet,
  Devoret, and Huard}}]{flurin:2012}
\bibinfo{author}{\bibfnamefont{E.}~\bibnamefont{Flurin}},
  \bibinfo{author}{\bibfnamefont{N.}~\bibnamefont{Roch}},
  \bibinfo{author}{\bibfnamefont{F.}~\bibnamefont{Mallet}},
  \bibinfo{author}{\bibfnamefont{M.~H.} \bibnamefont{Devoret}},
  \bibnamefont{and} \bibinfo{author}{\bibfnamefont{B.}~\bibnamefont{Huard}},
  \bibinfo{journal}{Phys. Rev. Lett.} \textbf{\bibinfo{volume}{109}},
  \bibinfo{pages}{183901} (\bibinfo{year}{2012}).

\bibitem[{\citenamefont{Ludwig et~al.}(2012)\citenamefont{Ludwig,
  Safavi-Naeini, Painter, and Marquardt}}]{ludwig:2012}
\bibinfo{author}{\bibfnamefont{M.}~\bibnamefont{Ludwig}},
  \bibinfo{author}{\bibfnamefont{A.~H.} \bibnamefont{Safavi-Naeini}},
  \bibinfo{author}{\bibfnamefont{O.}~\bibnamefont{Painter}}, \bibnamefont{and}
  \bibinfo{author}{\bibfnamefont{F.}~\bibnamefont{Marquardt}},
  \bibinfo{journal}{Phys. Rev. Lett} \textbf{\bibinfo{volume}{109}},
  \bibinfo{pages}{063601} (\bibinfo{year}{2012}).

\bibitem[{\citenamefont{Johansson et~al.}(2012)\citenamefont{Johansson, Nation,
  and Nori}}]{johansson:2012}
\bibinfo{author}{\bibfnamefont{J.~R.} \bibnamefont{Johansson}},
  \bibinfo{author}{\bibfnamefont{P.~D.} \bibnamefont{Nation}},
  \bibnamefont{and} \bibinfo{author}{\bibfnamefont{F.}~\bibnamefont{Nori}},
  \bibinfo{journal}{Comp. Phys. Comm.} \textbf{\bibinfo{volume}{183}},
  \bibinfo{pages}{1760 } (\bibinfo{year}{2012}).

\bibitem[{\citenamefont{Johansson et~al.}(2013)\citenamefont{Johansson, Nation,
  and Nori}}]{johansson:2013}
\bibinfo{author}{\bibfnamefont{J.~R.} \bibnamefont{Johansson}},
  \bibinfo{author}{\bibfnamefont{P.~D.} \bibnamefont{Nation}},
  \bibnamefont{and} \bibinfo{author}{\bibfnamefont{F.}~\bibnamefont{Nori}},
  \bibinfo{journal}{Comp. Phys. Comm.} \textbf{\bibinfo{volume}{184}},
  \bibinfo{pages}{1234 } (\bibinfo{year}{2013}).

\bibitem[{fig(2014)}]{figshare}
\bibinfo{journal}{The source code for the numerical simulations is available on
  Figshare, http://dx.doi.org/10.6084/m9.figshare.936910}
  (\bibinfo{year}{2014}).

\end{thebibliography}

\end{document}